\documentclass[pdflatex,sn-basic,referee]{sn-jnl}

\usepackage{graphicx}%
\usepackage{multirow}%
\usepackage{amsmath,amssymb,amsfonts}%
\usepackage{mathrsfs}%
\usepackage[title]{appendix}%
\usepackage[dvipsnames]{xcolor}%
\usepackage{textcomp}%
\usepackage{booktabs}%
\usepackage[nolist]{acronym}
\usepackage{natbib}
\usepackage{verbatim}
\usepackage{anyfontsize}
\usepackage{array}
\usepackage{physics}
\usepackage{ar}
\usepackage{orcidlink}

\newcommand{\mps}{~m~s$^{-1}$} 

\begin{document}

\title[Article Title]{Passive control of wing tip vortices through a grooved-tip design}

\author[1,3]{\fnm{Junchen} \sur{Tan}\email{jt988@cam.ac.uk}\orcidlink{0000-0001-5831-8538}}
\author[2]{\fnm{Sh\=uji} \sur{\=Otomo}\email{otomo@go.tuat.ac.jp}\orcidlink{0000-0002-0344-3961}}
\author[1]{\fnm{Ignazio Maria} \sur{Viola}\email{i.m.viola@ed.ac.uk}\orcidlink{0000-0002-3831-8423}}
\author*[1,3]{\fnm{Yabin} \sur{Liu}\email{yl742@cam.ac.uk}\orcidlink{0000-0002-0150-5671}}

\affil[1]{\orgdiv{School of Engineering, Institute for Energy Systems}, \orgname{University of Edinburgh}, \orgaddress{\city{Edinburgh}, \postcode{EH9 3FB}, \country{United Kingdom}}}
\affil[2]{\orgdiv{Department of Mechanical Systems Engineering}, \orgname{Tokyo University of Agriculture and Technology}, \orgaddress{\city{Koganei, Tokyo}, \postcode{184-8588}, \country{Japan}}}
\affil*[3]{\orgdiv{Department of Engineering}, \orgname{University of Cambridge}, \orgaddress{\city{Cambridge}, \postcode{CB2 1PZ}, \country{United Kingdom}}}


\abstract{
This paper investigates the characteristics and control of tip vortices generated by a finite wing, focusing on the impact of the novel grooved-tip designs.
Tip vortices can lead to flow loss, noise, vibration and cavitation in hydrodynamic systems.
We propose and develop a grooved-tip design, featuring multiple grooves distributed along the wing tip to alter the tip vortex structure and dynamics.
Four grooved-tip designs, including tilted and shrinking grooves, were experimentally investigated.
Streamwise and cross-flow Particle Image Velocimetry (PIV) measurements were employed to visualise the flow fields near the wing tip and along the primary tip vortex trajectory.
The PIV results demonstrate that the grooved-tip designs significantly reduce the velocity magnitude within the primary tip vortex.
This velocity deficit is attributed to the decreased suction within the vortex core.
Furthermore, cross-flow PIV measurements reveal that the tip separation vortex is substantially suppressed, and the strength of the primary tip vortex is significantly mitigated.
Downstream of the wing, the grooved tips lead to a reduction in vortex swirling strength and an enlargement of the vortex dimensions, suggesting enhanced diffusion and a reduction of the pressure drop of approximately 40\%.
Our findings highlight the potential of these grooved-tip designs to effectively modify tip vortex behaviour and mitigate the pressure drop within the tip vortex region, with negligible changes to the lift and drag performance.
This work can inform advanced passive vortex control strategies in wing- and blade-based systems, with potential applications in hydrofoils of marine vessels and underwater vehicles, as well as in turbines, propellers, pumps, etc.
\\
\begin{center}
    \textbf{Graphical Abstract}\\
    \includegraphics[width=\textwidth]{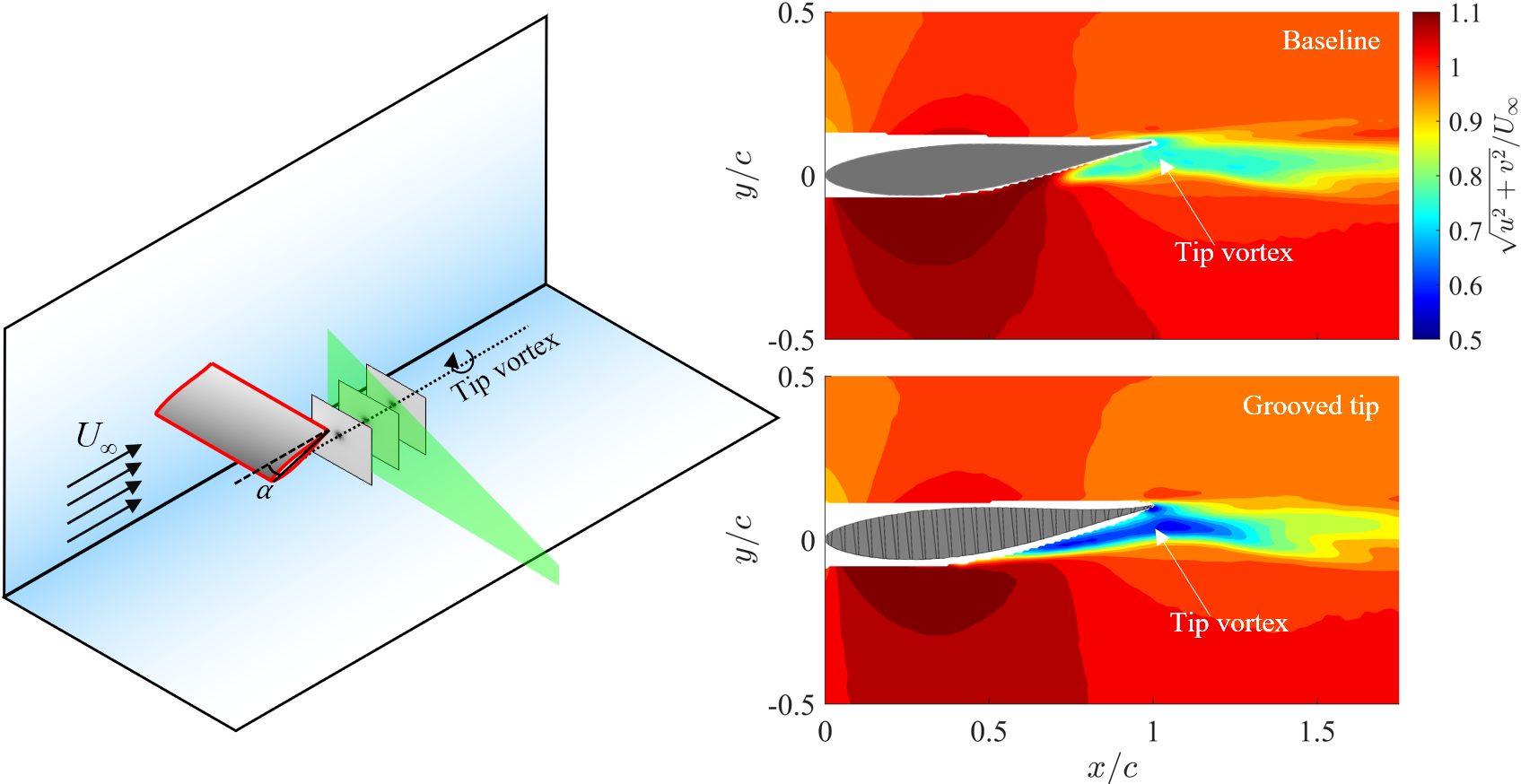}
\end{center}
}

\keywords{Wing, Tip vortices, Permeability, Groove}

\maketitle

\begin{acronym}
    \acro{tlv}[TLV]{tip leakage vortex}
    \acro{ptlv}[PTLV]{primary tip leakage vortex}
    \acro{ptv}[PTV]{primary tip vortex}
    \acro{tsv}[TSV]{tip separation vortex}
    \acro{tsr}[TSR]{tip speed ratio}
    \acro{ps}[PS]{pressure side}
    \acro{ss}[SS]{suction side}
    \acro{piv}[PIV]{Particle Image Velocimetry}
    \acro{cfd}[CFD]{Computational Fluids Dynamics}
    \acro{2d}[2D]{two-dimensional}
    \acro{piv}[PIV]{Particle Image Velocimetry}
    \acro{rms}[r.m.s.]{root-mean-square}
    \acro{tev}[TEV]{trailing edge vortex}
    \acro{adv}[ADV]{Acoustic Doppler Velocimetry}
    \acro{fsl}[FSL]{full-scale load}
    \acro{rans}[RANS]{Reynolds-averaged Navier–Stokes}
\end{acronym}

\section*{List of selected symbols}
\begin{tabular}{ll}
    $s\AR$                                                  & Semi-aspect ratio\\
    $c$                                                     & Chord length\\
    $C_D$                                                   & Drag coefficient\\
    $C_L$                                                   & Lift coefficient\\
    $d$                                                     & Diameter\\
    O($x,\,y,\,z$)                                          & Cartesian coordinates: streamwise, cross-stream, spanwise\\
    O($r,\,\theta,\,X$)                                     & Cylindrical coordinates: radial, azimuthal, axial\\
    $p$                                                     & Pressure\\
    $r_c$                                                   & Viscous core radius\\
    $Re_c$                                                  & Chord-based Reynolds number\\
    $\vb*{u} = (u, \,v, \,w)$                               & Fluid velocity\\ 
    $U_\infty$                                              & Free stream velocity\\
    $\alpha$                                                & Angle of attack\\
    $\Gamma$                                                & Circulation\\
    $\lambda_c$                                             & Swirling strength\\
    $\rho$                                                  & Fluid density\\
    $\omega$                                                & Vorticity\\
\end{tabular}

\section*{Abbreviations}
\begin{tabular}{ll}
    PIV     & Particle image velocimetry\\
    PTLV    & Primary tip leakage vortex\\
    PTV     & Primary tip vortex\\
    TEV     & Trailing edge vortex\\
    TLV     & Tip leakage vortex\\
    TSV     & Tip separation vortex\\
\end{tabular}

\section{Introduction}\label{intro}
The formation of tip vortices, driven by the pressure difference between the pressure and suction sides of a finite wing or blade tip, generate a swirling flow in the wake.
These vortices and associated problems, such as induced drag, wakes and noise, have been a long-standing concern in wing- and blade-based systems over the past decades. 
From an aerodynamic point of view, wing tip vortices contribute to induced drag, accounting for up to 40\% of the total drag \citep{margaris2006}. 
Furthermore, tip leakage vortices are responsible for nearly one-third of the aerodynamic loss in gas turbine cascades \citep{denton1993}. 
In hydrodynamic applications such as propellers, pumps and tidal turbines, tip vortices and tip leakage vortices are often linked to cavitation due to the pressure drops within the vortex core \citep{pereira2004,liu2018}, which is detrimental due to increased noise \citep{wittekind2016,pennings2016}, structural vibrations \citep{cheng2025} and reduces efficiency \citep{aktas2016,capone2023}.
Additionally, applications in marine environment face a higher risk of cavitation erosion due to the corrosive nature of seawater \citep{hou2014}.
Hydrofoils are increasingly used in sailing \citep{ng2025} and commercial vessels \citep{godo2023} for their efficiency and performance benefits, though cavitation remains a key challenge in their design and operations \citep{godo2023b,godo2024}.

With the growing emphasis on renewable energy, tidal stream power is expected to play a vital role in achieving the net-zero transition by 2050. 
Tidal turbines, however, also face persistent challenges related to tip vortices and cavitation, limiting their efficiency and reliability when operating at higher \ac{tsr} \citep{adcock2021fluid}.
Attenuating tip vortices can lead to enhanced power output and reduced mechanical stress on powertrain components \citep{ning2014understanding,wimshurst2018}.
Moreover, the efficiency of tidal and wind turbines is often constrained by turbine-to-turbine wake interactions, which can result in power losses of up to 20\% \citep{barthelmie2011}. A key contributor to those losses is tip vortices, which significantly influences the generation and propagation of the turbine wake \citep{fischereit2022}.
Reducing the intensity of the tip vortex, such as by lowering the axial-to-circumferential momentum ratio of the swirling vortex, can trigger earlier vortex breakdown \citep{robinson1994}.
This earlier breakdown has been shown to facilitate faster wake recovery \citep{lignarolo2015}, which is beneficial for improving turbine array performance.
Therefore, effective control of tip vortices and associated pressure drops and wakes is critical for the further development of the tidal energy sector. 
Given the widespread occurrence of vortex-induced cavitation across various underwater applications, including ship propellers, underwater vehicles, tidal stream turbines, hydropower turbines and pumps, developing advanced approaches for controlling tip vortices and mitigating the cavitation risks can also inform novel technologies in transformative engineering applications.

In essence, \ac{ptlv}, which occurs in the presence of a small gap between the blade tips and the casing wall, and \ac{ptv}, which forms in the absence of a casing wall, share the same underlying flow physics.
Due to the pressure difference between the pressure and suction sides of a wing or a blade, the flow near the tip rolls up into a swirling vortex, resulting in a low-pressure region at its core \citep{liu2024arxiv,bi2024}.
Various active and passive strategies have been proposed to mitigate the \ac{ptlv} and \ac{ptv}. 

Active strategies to control tip vortices mainly focused on wing-based systems, such as oscillating control surfaces to vary the degree of boundary layer separation by means of zero ``mass-flux perturbations'' \citep{greenblatt2005}, synthetic jet to produce vortex diffusion and stretching \citep{margaris2006,dghim2020,zaccara2022} and plasma actuation to displace the tip vortices \citep{dong2022}.
For helicopter rotor blades, active twist actuation has demonstrated the ability to control both the strength and trajectory of the blade tip vortices \citep{bauknecht2017}.
However, these active strategies substantially introduce additional structural and system complexity and require external power.

Passive strategies primarily include geometric modification to the tip region.
For example, \citet{ji2021} numerically investigated the thickened and raked blade tips for a pump-jet propulsor, reporting a decrease in the \ac{tlv} strength and significantly improved negative pressure peaks; however, the rotor efficiency decreased due to reduced thrust and lift-to-drag ratio at outer radii.
\citet{wang2023} found that a wavy blade tip divided the single large-scale \ac{ptlv} into smaller-scale vortices along the tip chord, reducing the \ac{ptlv} swirling strength but increasing the swirling near the suction side due to secondary tip leakage vortices.

One of the most promising passive approaches to control tip vortices is the C-shaped single-groove design by \citet{liu2018b}.
The groove generates a jet that impinges on the \ac{ptlv}, suppressing its development across a range of inflow velocities.
However, a follow-up study by \citet{jiang2022} experimentally observed that the shrinking C-shaped groove at the blade tip could induce extra cavitation due to high local velocity inside the groove.
A recent study by \citet{bi2024} demonstrated the potential of mitigating cavitation due to \ac{tlv} through the application of a thin layer of porous metal at the blade tip.
However, the study primarily analysed the \ac{tlv} behaviour across different tip gap sizes without examining the role of permeability.
More recently, \citet{liu2024arxiv} proposed a permeable tip treatment to tidal turbine blade tips, inspired by the findings of \citet{cummins2018} who demonstrated the role of permeability in controlling vortical structures in the separated vortex ring behind a dandelion seed.
Their analysis covered a range of permeability and suggested that an optimal permeability for controlling the tip vortices exists, which can most effectively mitigate the associated pressure drop through enhanced vortex diffusion and enlarged vortex dimensions.
In a similar vein, \citet{liu2025controlling} conducted a numerical study on a compressor cascade, demonstrating significant effects of permeable tip treatment in controlling \ac{tlv}. 

Building upon our previous work \citep{liu2018b, liu2020, liu2024arxiv, liu2025controlling}, this study aims to develop a novel passive approach that combines the benefits of groove-jet impingements and permeable tip treatment to mitigate the tip vortices and associated pressure drop, while achieving an effective \ac{2d} permeability.
Advancing from our earlier single-grove design \citep{liu2018b,liu2020,han2022method}, the present approach utilises multiple grooves distributed along the entire tip chord.
Unlike the single-groove design, which may introduce additional local pressure minima due to concentrated high-speed jet flow within the groove, the distributed multi-groove configuration approximates uniform permeability, thereby avoiding the undesirable effects.
Experiments were performed on a wing model in a water tunnel, with time-averaged flow fields on streamwise and cross-flow planes visualised using \ac{piv} measurements for a baseline case and four grooved-tip designs.
In this paper, we have demonstrated that the fundamentally novel concept of controlling tip vortices through tip permeability can be achieved by practical designs.

The rest of the paper is organised as follows.
The lift and drag performance of the baseline and modified wings are compared in Section~\ref{sec:3.1}.
In Section~\ref{sec:3.2}, we examine the velocity deficit within the tip vortices, highlighting the alleviation of pressure drop around the tip vortex core.
In Section~\ref{sec:3.3}, the vortex structures with and without the grooves are compared to elucidate changes in the flow patterns.
In Section~\ref{sec:3.4}, the change in the vortex swirling strength is discussed and an analytical relationship between the swirling strength and pressure drop is derived to estimate the pressure change with the tip vortex core.


\section{Methodology}\label{methods}
\subsection{Experimental setup}\label{methods:experimental}
\begin{figure*}[tbp!]
    \centering
    \includegraphics[width=7.23cm]{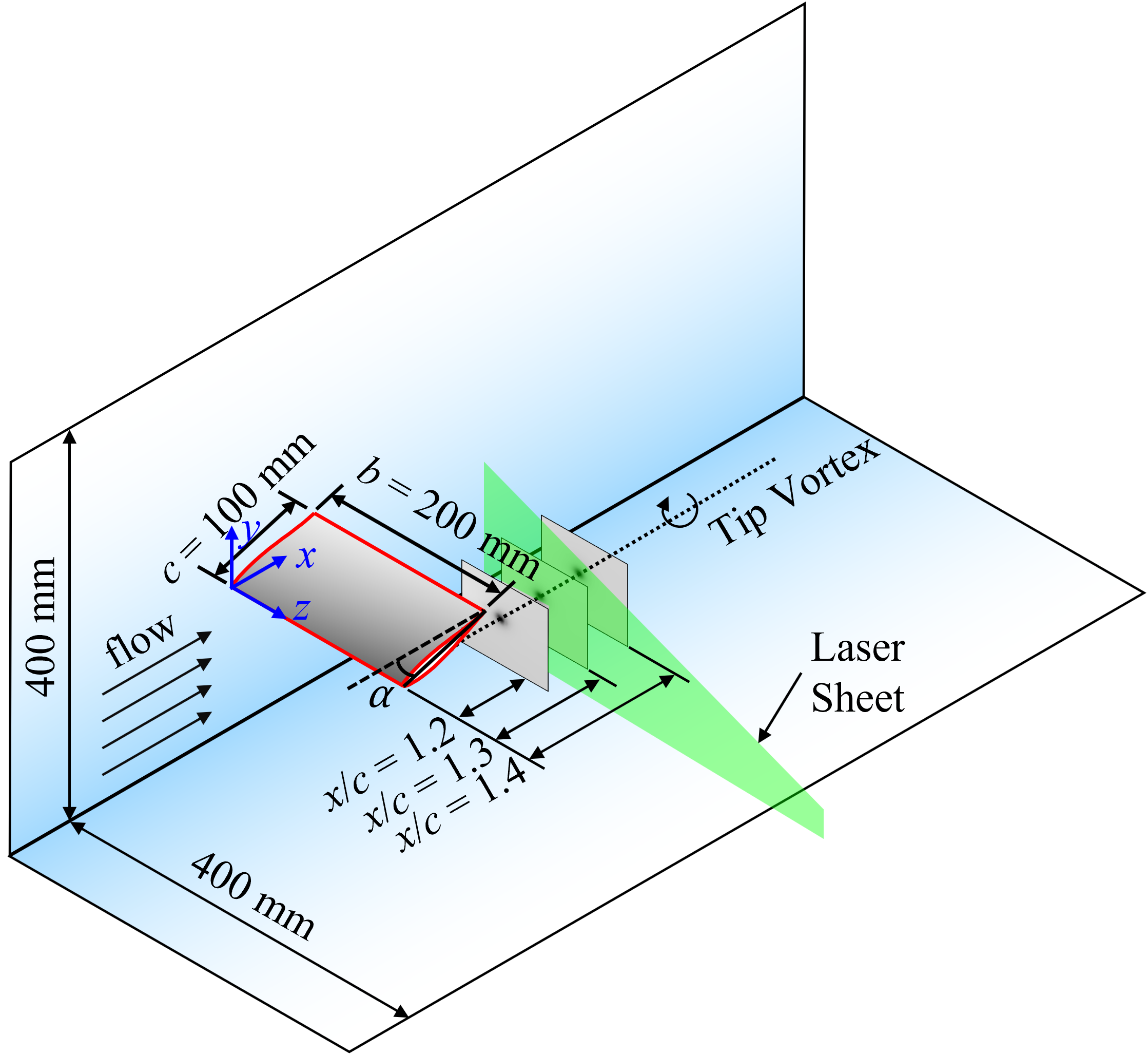}
    \caption{Dimensions of the water tunnel and coordinate system used in the present study. The displayed contours are for illustrative purposes.}
    \label{fig:coordinate_system}
\end{figure*}

The experiments were conducted in a water tunnel located at the University of Edinburgh \citep{robinson2015}. 
The water tunnel is 8~m long, 0.4~m wide, and filled with water to a depth of 0.4~m; flow is preconditioned by curving vanes before entering and exiting the channel.
The half-wing model has a thickened NACA¬63-415 hydrofoil profile, consistent with the blade tip geometry in the UK Supergen ORE Tidal Benchmarking Project \citep{harvey2023oxfordexperiment}.
The wing features a chord length of $c = 0.1$~m and a span of $b = 0.2$~m, resulting in a semi-aspect ratio ($b/c$) of $s\AR = 2$.
The coordinate system O$(x, \,y, \,z)$ indicates the streamwise, cross-stream and spanwise coordinates (see Figure~\ref{fig:coordinate_system}). The origin of the coordinate system is located at the leading edge at the wing root.
The velocity components in each direction are $u$, $v$ and $w$, respectively. 
The free stream velocity was calibrated using a Nortek Vectrino acoustic Doppler velocimeter at the centre of the measurement plane, in the absence of the wing and underwater cameras. 
Experiments were performed at a nominal velocity of $U_\infty = 0.3$\mps, equivalent to a Reynolds number based on the chord length of $Re_c=3.22 \times 10^4$, with a streamwise turbulence intensity of approximately 4.5\%.
Lift and drag were measured over a wide range of angles of attack, while for flow visualisations the angle of attack was set at $\alpha = 6^\circ$, which is close to the hydrofoil’s optimal lift-to-drag ratio \citep{harvey2023oxfordexperiment}.
At $\alpha = 6^\circ$, the leading edge was positioned 190~mm below the free water surface.

\subsection{Particle image velocimetry measurements}
\begin{figure*}[tbp!]
    \centering
    \includegraphics[width=\textwidth]{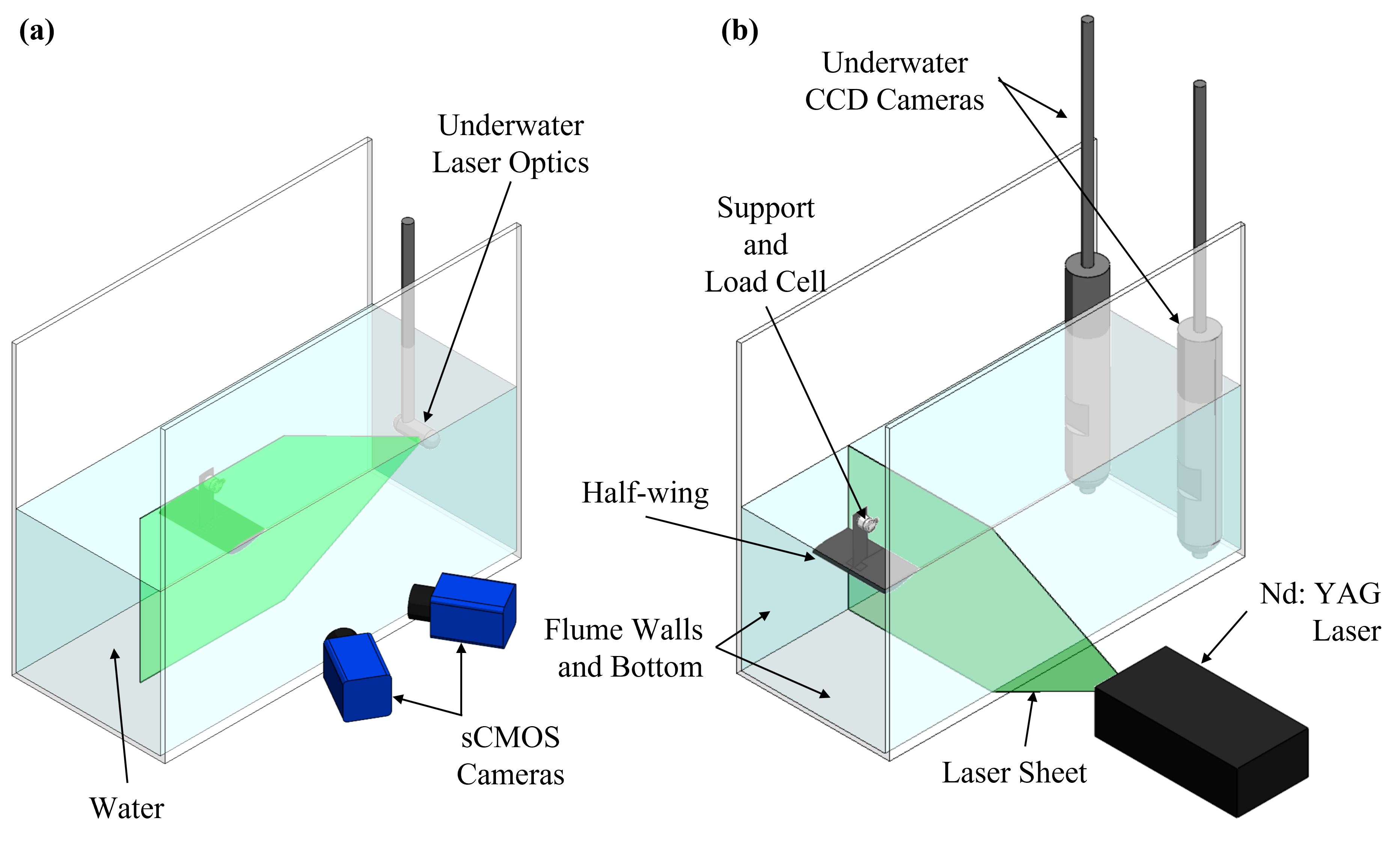}
    \caption{Experimental setup for the particle image velocimetry (PIV) measurements: (a) streamwise, (b) cross-flow.}
    \label{fig:experimental_setup}
\end{figure*}

A schematic diagram of the inverted half-wing model, along with the experimental setup for \ac{piv} measurements, is shown in Figure~\ref{fig:experimental_setup}.
Model inversion was employed to minimise the flow disturbance on the suction surface from the support structure and the load cell.
Quantitative flow field measurements were performed using a LaVision FlowMaster stereoscopic \ac{piv} (2D3C) system.
Schematic diagrams of the experimental setup for the streamwise (on $x-y$ planes) and cross-flow (normal to free stream, on $y-z$ planes) \ac{piv} measurements are shown in Figures~\ref{fig:experimental_setup}(a) and (b), respectively.
Streamwise measurements were conducted at a fixed plane slightly inboard of the tip, at $x/c = 1.95$, whereas the cross-flow measurement planes were evenly spaced between $x/c = 0.6 - 1.5$ with a $0.1c$ increment.

A NewWave Solo 200XT double-pulse Nd:YAG laser, with a maximum energy of 200~mJ/pulse was used to illuminate the desired planes.
Water was seeded with silver-coated hollow glass spheres with a mean diameter of 10~{\textmu}m and a density of 1.4~g~cm$^{-3}$.
\ac{piv} images were captured by two Imager sCMOS cameras placed on the side of the water tunnel for the streamwise measurements, or two Imager ProSX 5M cameras in sealed torpedo-shaped compartments placed downstream of the wing for cross-flow measurements.
Scheimpflug adapters were used to compensate for the oblique viewing angle from the perspective view. Cameras and laser pulses were triggered simultaneously by a PTU X synchroniser.
For each measurement, 500 instantaneous flow fields over 850 convecting periods ($c/U_\infty$) were captured at a rate of 1.75~Hz to accurately account for the mean flow statistics.
The commercial software package LaVision DaVis~10.2 was used to calculate the velocity vectors.
The interrogation window size was $64 \times 64$ pixels, which was then reduced to $32 \times 32$ pixels with a 75\% overlap.
The effective grid size was approximately 1\%$c$.
The uncertainty of the velocity measurements was reported by the DaVis 10.2 software to be within 2\% of the mean free stream velocity, where the uncertainties in the 3D velocity components $(u, v, w)$ are calculated by propagating the known uncertainties of the \ac{2d} displacements from each camera, namely $(u_1, v_1)$ and $(u_2,v_2)$, through the calibration mapping function \citep{davis}.

It is important to note that the underwater camera casings have a diameter of 89~mm and span across the entire water depth, which unavoidably introduced blockage to the flow (see Figure~\ref{fig:experimental_setup}(b)).
Consequently, the measured free stream velocity in the presence of the underwater \ac{piv} cameras was reduced to $U_\infty = 0.28$\mps, corresponding to an updated chord-based Reynolds number of $Re_c=2.80\times10^4$ for cross-flow measurements.
A similar underwater flow visualisation technique was previously used in our laboratory \citep{muir2017underwater}.

\subsection{Force measurements}
Lift and drag measurements were conducted over a range of angles of attack from $\alpha = 0^\circ - 14^\circ$, in increments of $1^\circ$.
The chordwise ($F_x$) and chord-normal ($F_y$) forces were collected using an ATI Industrial Automation Nano17 IP68 six-axis load cell with a \ac{fsl} of 25~N and a resolution of 0.00625~N, and logged to a computer running the ATIDAQ .Net software through a National Instruments USB-6210 data acquisition device.
Force data were recorded with a sampling frequency 1000~Hz.
For each measurement, force signals were recorded for 120~s over 360 convecting periods ($c/U_\infty$), which is sufficiently long for the mean and \ac{rms} of the signals to converge.
Buoyancy forces and load cell factory offset were accounted for in the water tunnel for all measured angles of attack; the load cell was zeroed in quiescent flow before each measurement to obtain the hydrodynamic forces.
Axis rotation was performed to convert $F_x$ and $F_y$ into the hydrodynamic lift and drag.
The maximum uncertainty in the force measurements are estimated to be approximately $B_{T_\mathrm{max}}(C_L) = \pm 0.090$ and $B_{T_\mathrm{max}}(C_D) = \pm 0.030$, respectively, based on the methods of \citet{moffat1985}.
The relatively large uncertainties are attributed to the fact that the measured force was a small percentage of the \ac{fsl}.
For example, the maximum recorded lift was 2.65\% of the \ac{fsl}, while the maximum recorded drag was only 0.41\% of the \ac{fsl}.
The quantification of force measurements uncertainties are detailed in Appendix~\ref{app:uncertainty}.

\subsection{Models with grooved tips}\label{methods:groovedtips}
\begin{figure*}[tbp!]
    \centering
    \includegraphics[width=12.32cm]{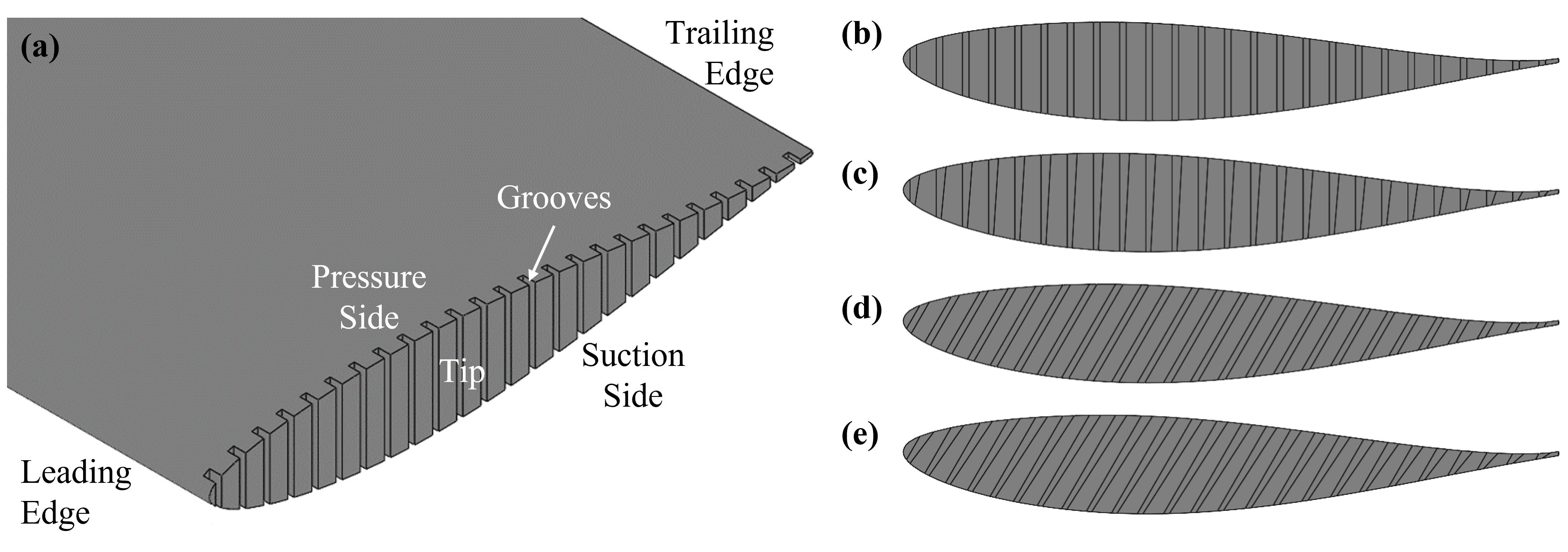}
    \caption{NACA 63-415 hydrofoils with grooved tip: (a) and (b) 25G-R-CN, (c) 25G-C-CN, (d) 25G-R-P, (e) 25G-C-P.}
    \label{fig:wing_grooves}
\end{figure*}

\citet{liu2024arxiv} suggests that an optimal range of tip permeability exist for suppressing the pressure drop due to tip vortices for a model-scale tidal turbine.
The optimal mitigation effect is achieved when the non-dimensional permeability, Darcy number $Da$ $=\kappa / \bar{\tau}^2$ (here, $\kappa$ is the permeability and $\bar{\tau} = 0.0952c$ is the average tip thickness), falls between $10^{-6}$ and $10^{-5}$.
Accordingly, we propose multiple-groove configurations, with twenty-five evenly distributed grooves. Each groove having a nominal width of $0.01c$ (in the chordwise direction) and a fixed depth of $0.02c$ (in the spanwise direction), designed to produce an equivalent \ac{2d} permeability within this range.
For the determination of the \ac{2d} permeability of the grooved tips, please refer to Appendix~\ref{app:2dpermeability}.
Four different groove geometries are investigated, as illustrated in Figure~\ref{fig:wing_grooves}.
These designs vary in terms of the contraction ratios from the \ac{ps} to the \ac{ss} and the orientation for each groove.
Overall, five half-wing models were manufactured, comprising one baseline case without any tip modification and four grooved-tip designs, as shown in Figures~\ref{fig:wing_grooves}(b)--(e).
The numerically-derived chord-normal non-dimensional permeability $Da$ for the four grooved-tip designs are summarised in Table~\ref{tab:permeability}.

\begin{table}[tbp!]
    \centering
    \caption{Comparison of $Da$ values for different tip designs.}
    \label{tab:permeability}
    \begin{tabular}{m{15em} m{7.5em}}
        \hline
        \textbf{Tip configuration}  & \textbf{$Da$} ($\times10^{-5}$) \\
        \hline
        Baseline                    & $--$  \\
        25G-R-CN                    & 2.20 \\
        25G-C-CN                    & 1.70 \\
        25G-R-P                     & 1.59 \\
        25G-C-P                     & 1.19 \\
        \hline
    \end{tabular}
\end{table}


\section{Results and discussion}\label{results}
\subsection{Lift and drag performance}\label{sec:3.1}
\begin{figure*}[tbp!]
    \centering
    \includegraphics[width=5.60cm]{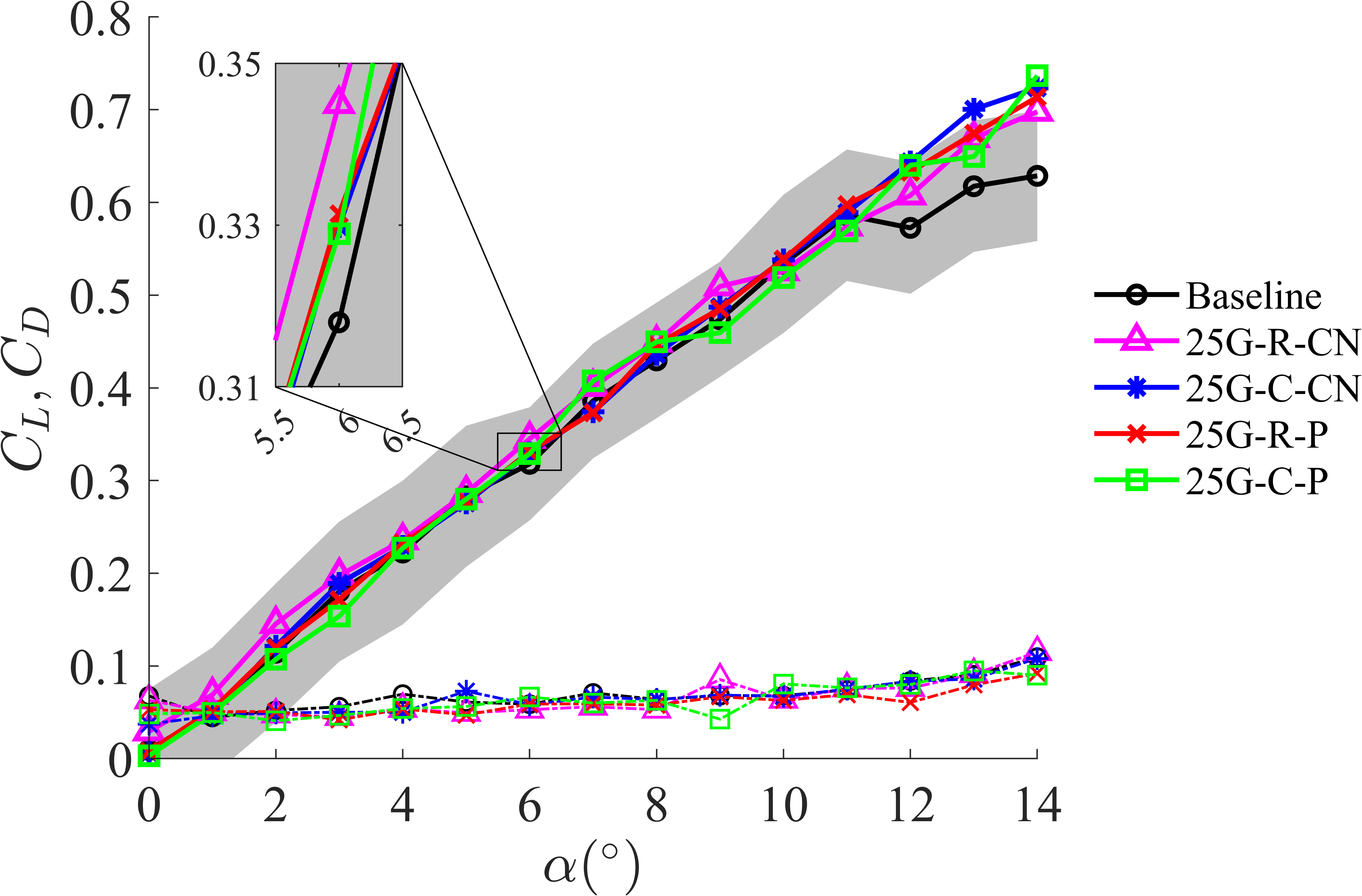}
    \caption{Variations of the lift coefficient $C_L$ (solid lines) and drag coefficient $C_D$ (dashed line) as a function of angle of attack $\alpha$. The shaded region indicates the interval within one standard deviation from the mean $C_L$ of the baseline case.}
    \label{fig:line_CL_CD}
\end{figure*}

This section evaluates the influence of the grooved-tip design on the aerodynamic performance of the wing, alongside primary objective of controlling tip vortices and mitigating the pressure drops within the \ac{ptv} rigion.
Figure~\ref{fig:line_CL_CD} presents the variations of the lift ($C_L$) and drag ($C_D$) coefficients as a function of the angle of attack ($\alpha$) for the baseline and grooved-tip designs.
The lift coefficients (solid lines) of the baseline and all grooved-tip designs are found to be comparable, particularly within the pre-stall angles of attack that all mean $C_L$ values for the grooved-tip designs fall within one standard deviation of the baseline case indicated by the shaded region in Figure~\ref{fig:line_CL_CD}.
A moderate delay in onset of stall is observed at higher angles of attack for wings with grooved tips, suggesting that these designs may contribute to delaying flow separation in the near-tip region.
The drag coefficients (dashed lines)  generally follow a similar trend, although they exhibit a greater degree of scattering due to the relatively higher percentage uncertainty of drag measurements.

The inset in Figure~\ref{fig:line_CL_CD} highlights the differences in the lift coefficient between the baseline case and grooved-tip designs at $\alpha = 6^\circ$.
At this specific angle of attack, the grooved tips yield slightly higher lift coefficients than the baseline case; however, these differences are relatively small and fall within the range of measurement uncertainties.
Overall, these results indicate that the grooved-tip designs have a positive, albeit limited, impact on the wing's lift and drag performance.

For the current experiments, the grooves at the wing tip have a depth of 2\% of the tip chord length and 1\% of the wing span.
It is, therefore, anticipated that when such a grooved-tip design is scaled for application in systems like tidal turbine blades, where the grooved-tip structure contributes a much smaller proportion of the turbine diameter (e.g., 0.1\% of the turbine diameter in \cite{liu2024arxiv}), the grooved-tip design is unlikely to result in a significant change in the overall power output or thrust in normal operating \ac{tsr}.

\subsection{Velocity deficit within the primary tip vortex}\label{sec:3.2}
\subsubsection{Velocity magnitude deficit}
\begin{figure*}[tbp!]
    \centering
    \includegraphics[width=9.528cm]{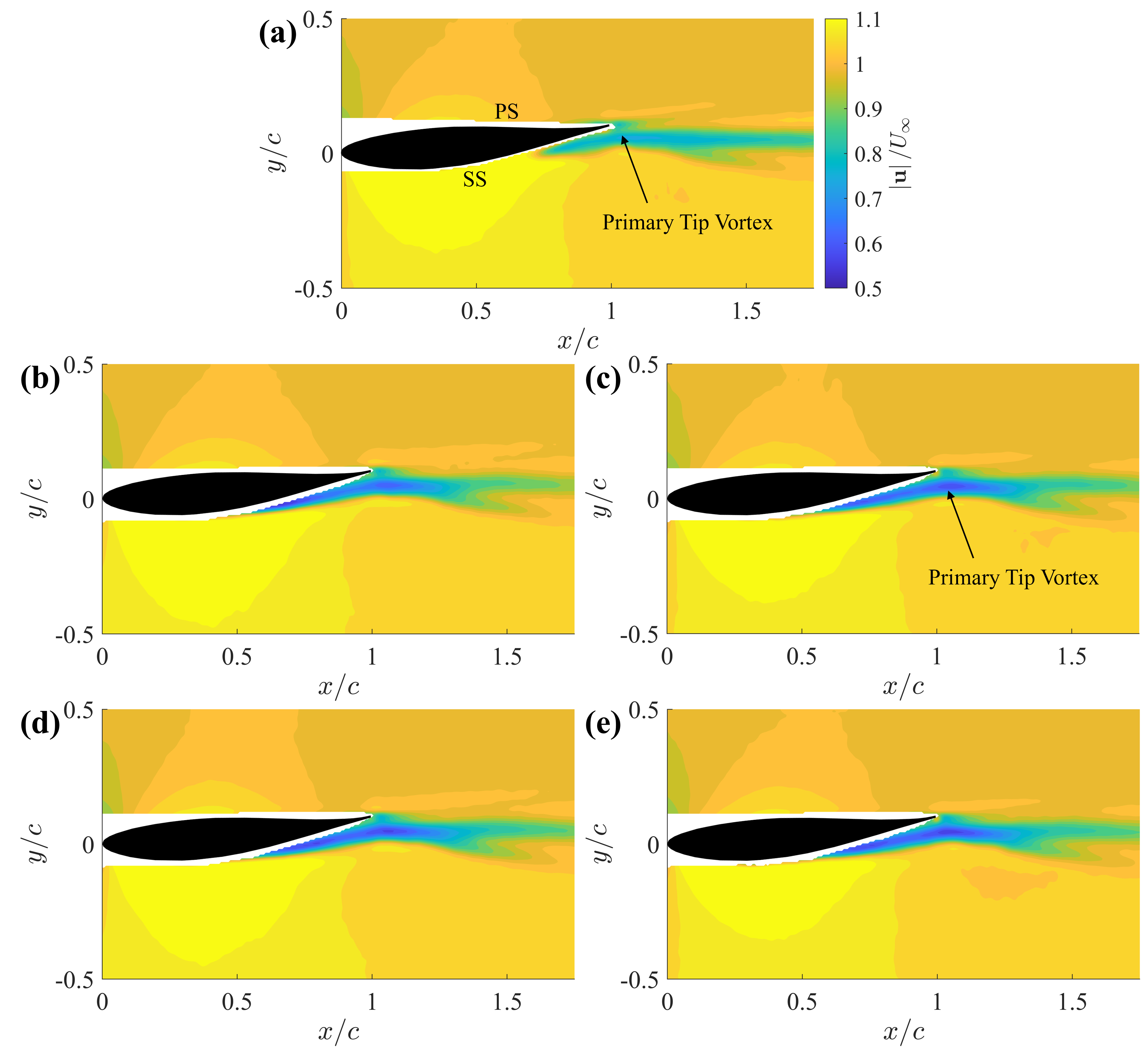}
    \caption{Time-averaged velocity magnitude contours at $z/c = 1.95$ streamwise plane for (a) baseline, (b) 25G-R-CN, (c) 25G-C-CN, (d) 25G-R-P, (e) 25G-C-P. SS: suction side; PS: pressure side.}
    \label{fig:streamwisepiv_Vmag}
\end{figure*}

This section presents the velocity field results obtained from the \ac{piv} measurements, primarily demonstrating how grooved-tip designs effectively reduce the velocity magnitude within the \ac{ptv} region, leading to a significant alleviation of the associated pressure drop.
Figure~\ref{fig:streamwisepiv_Vmag} shows the time-averaged velocity magnitude contours on a streamwise plane at $z/c=1.95$ (see Figure~\ref{fig:experimental_setup}(a) for experimental setup) for both the baseline case the four different grooved-tip designs.
While the free stream velocities remain constant across all cases due to repeatable inflow conditions, a notable velocity deficit is observed within the \ac{ptv} region for the grooved tips.
For instance, the minimum velocity magnitude within the \ac{ptv} significantly reduced from 0.76\mps for the baseline case (Figure~\ref{fig:streamwisepiv_Vmag}(a)) to 0.60\mps for the 25G-C-CN tip (Figure~\ref{fig:streamwisepiv_Vmag}(c)), representing a 21\% reduction.
This observation is meaningful as it suggests that the grooved-tip design can substantially alleviate the pressure drop inside the tip vortex.
This inference aligns with the pressure distribution of a steady axisymmetric vortex based on the governing equations \citep{batchelor1964}, which is qualitatively valid for the low-to-moderate Reynolds number (in the present study, $Re_c=2.80\times10^4$) and incompressible flow conditions in the water tunnel, where inertial effects are dominant and viscous dissipation is moderate.
In addition, \citet{zhao2024} verified the pressure-velocity relation, $\bar{p}\sim\frac{1}{2}\bar{u}^2$, derived from simplified \ac{rans} equation using three-dimensional experimental data for a tip vortex.
The derived relation signifies that the mean static pressure inside a tip vortex is governed by the local mean axial velocity, resembling the relationship described by Bernoulli’s principle.

\begin{figure*}[tbp!]
    \centering
    \includegraphics[width=11.544cm]{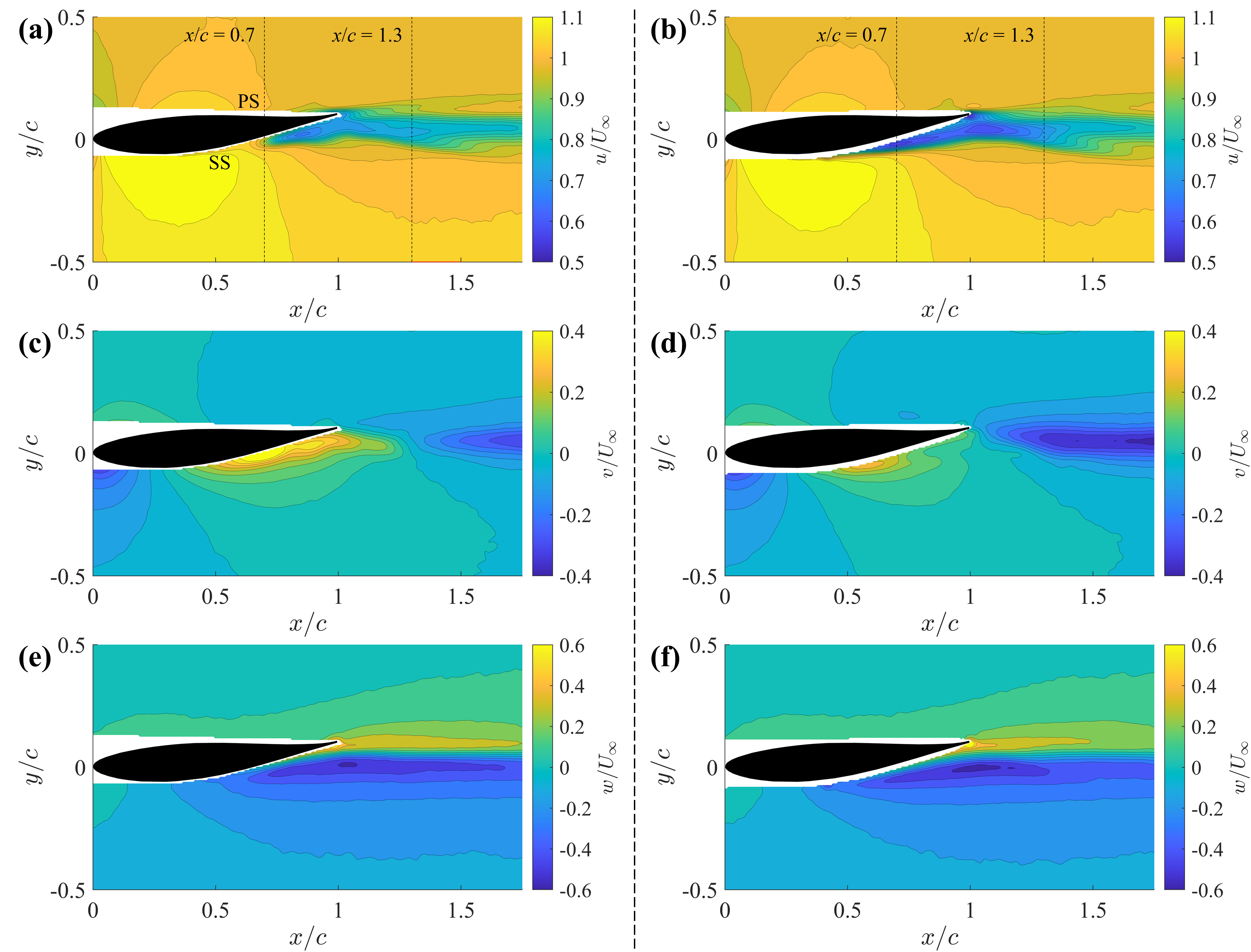}
    \caption{Time-averaged streamwise (a, b), cross-stream (c, d), and spanwise (e, f) velocity contours at $z/c = 1.95$ plane for baseline (a, c, e) and 25G-C-CN (b, d, f).}
    \label{fig:streamwisepiv_uvw}
\end{figure*}

An in-depth assessment of the individual velocity component, including the streamwise ($u/U_\infty$), cross-stream ($v/U_\infty$) and spanwise ($w/U_\infty$) velocities, on the streamwise plane $z/c=1.95$ as shown in Figure~\ref{fig:streamwisepiv_uvw} for the baseline case and 25G-C-CN design as an example.
Figure~\ref{fig:streamwisepiv_uvw} reveals that the velocity magnitude deficit in the \ac{ptv} region is primarily dominated by the streamwise velocity change.
In particular, the noticeable reduction in streamwise velocity within the \ac{ptv} region is similar to the observation in the velocity magnitude contours in Figure~\ref{fig:streamwisepiv_Vmag}.

As shown in the top row of Figure~\ref{fig:streamwisepiv_uvw}, the streamwise velocity contours illustrate a larger region of relatively high velocity near the trailing edge for the baseline case (Figure~\ref{fig:streamwisepiv_uvw}(a)), which is significantly reduced for the 25G-C-CN design (Figure~\ref{fig:streamwisepiv_uvw}(b)).
This reduction in streamwise velocity within the \ac{ptv} region is critical as it is a direct implication that the grooved-tip design exhibits a relatively higher local static pressure and a reduced pressure drop, according to the pressure-velocity relation derived by \citet{zhao2024}.
Furthermore, for the grooved-tip design, the more concentrated low-velocity region (indicated by the blue colour) shifts upstream, suggesting the groove-jet impinges on and mixes with the \ac{ptv}.

The cross-stream velocity contour further highlight the impact of the grooved tip.
Figure~\ref{fig:streamwisepiv_uvw}(c) for the baseline case shows a steeper velocity gradient on the suction side compared to the 25G-C-CN case in Figure~\ref{fig:streamwisepiv_uvw}(d), where the grooves effectively reduce this velocity gradient.
This reduction in pressure gradient on the suction side  may contribute to delaying boundary layer separation on wings with grooves.
Additionally, the upward cross-stream velocity region near the suction side extends further downstream for the baseline case, beyond the trailing edge, due to the sweeping effects from the \ac{tsv}.
In contrast, the 25G-C-CN case exhibits a strong downward wake immediately after leaving the trailing edge (Figure~\ref{fig:streamwisepiv_uvw}(d)), indicating reduced impact from the \ac{tsv} and enhanced mixing of the wake with the free stream.
In the bottom row, while the out-plane (yellow) velocities are comparable between the two cases, the 25G-C-CN case (Figure~\ref{fig:streamwisepiv_uvw}(f)) shows the in-plane (blue) velocity extending further from the wing surface compared to the baseline (Figure~\ref{fig:streamwisepiv_uvw}(e)), implying a more diffused tip vortex structure.
These detailed velocity components analysis provide further evidence of the effectiveness of the grooved-tip design in modifying the flow field around the wing tip.

\subsubsection{Axial velocity deficit}
\begin{figure*}[tbp!]
    \centering
    \includegraphics[width=8.5cm]{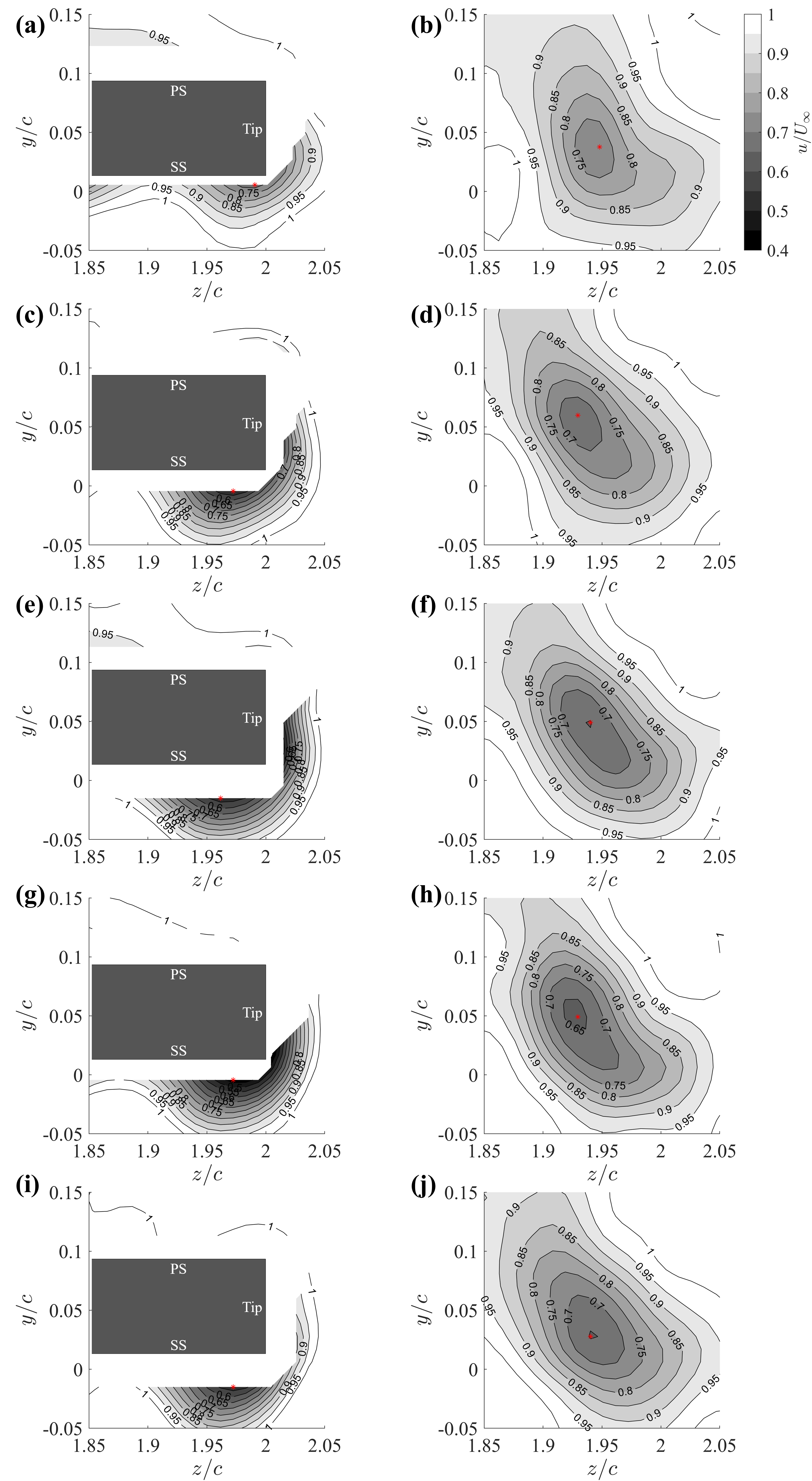}
    \caption{Time-averaged streamwise velocity contours at $x/c = 0.7$ (left column) and $x/c=1.3$ (right column) cross-flow plane for (a, b) baseline, (c, d) 25G-R-CN, (e, f) 25G-C-CN, (g, h) 25G-R-P, (i, j) 25G-C-P. The red asterisk (*) indicates the location of minimum velocity.}
    \label{fig:axialvelocity}
\end{figure*}

Figure~\ref{fig:axialvelocity} provides a comparison of the time-averaged streamwise velocity contours at two cross-sections of the tip vortices, namely $x/c=0.7$ (left column, on-wing) and $x/c=1.3$ (right column, in-wake), for the baseline and four grooved-tip configurations. In these plots, darker shades correspond to lower velocities, indicating a greater velocity deficit, while lighter colour represent velocities closer to the free stream velocity.

For the baseline case (Figures~\ref{fig:axialvelocity}(a) and (b)), a prominent region of velocity deficit is observed in the \ac{ptv} region near the wing tip, with the vortex core reaching a minimum value of approximately $u/U_\infty=0.75$.
In contrast, all grooved-tip designs exhibit noticeable changes compared to the baseline.
Notably, a more significant streamwise velocity deficit is consistently observed in the vortex cores during the tip vortex roll-up process (left column) and after the vortex is fully developed (right column).
The spanwise stretching of the low-velocity region at $x/c=0.78$ suggests an earlier onset of vortex roll-up and increased entrainment, potentially due to the enhanced mixing between the groove-jets and the \ac{ptv}.
Assuming the vortex axis is almost parallel with the free stream, the streamwise velocity component $u$ serves as a reasonable proxy for the vortex axial velocity $v_X$.
Consequently, the observed reduction in the streamwise velocity provides further evidence that grooved-tip designs are effective in mitigating the pressure drop with the tip vortex.

\begin{figure*}[tbp!]
    \centering
    \includegraphics[width=4.72cm]{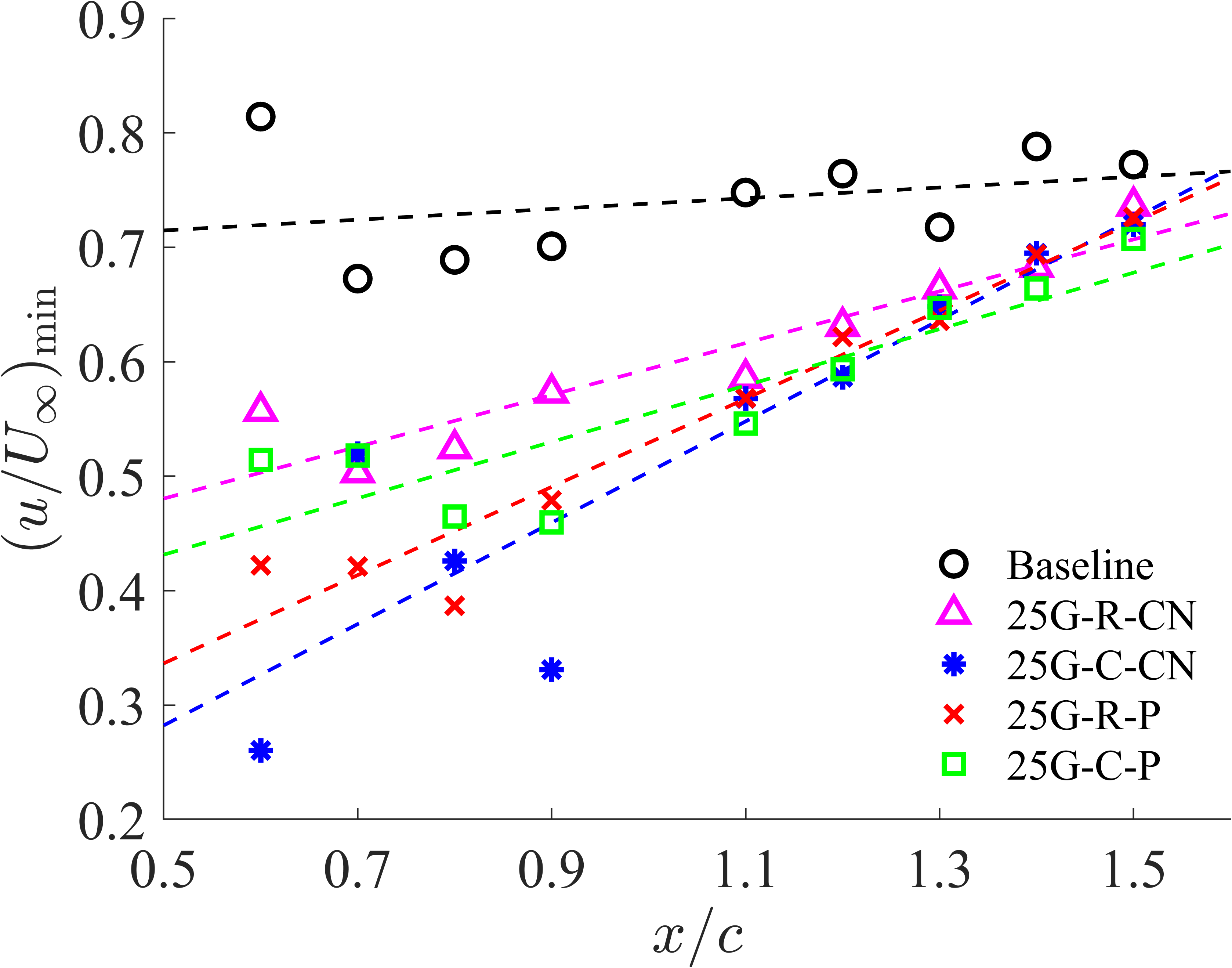}
    \caption{Comparison of streamwise velocity on each cross-section along the vortex core trajectory between baseline and grooved tips.}
    \label{fig:line_vel_mag}
\end{figure*}

To further analyse this phenomenon and account for the \ac{ptv}'s trajectory, we extract and present the minimum streamwise velocity magnitude along the tip vortex trajectory on all cross-flow planes, ranging from near the mid-chord ($x/c=0.6$) to half chord-length downstream of the trailing edge ($x/c=1.5$), as shown in Figure~\ref{fig:line_vel_mag}.
Particular attentions were paid to ensure the minimum velocity was extracted within the \ac{ptv} region.
The velocity profiles for all configurations consistently show an increasing trend downstream, signifying the expected diffusion and decay of tip vortices as they convect downstream from the wing tip.
Compared to the baseline case, the grooved-tip designs generally exhibit lower minimum streamwise velocity, despite some data scattering is observed.
These results, again, reinforces that grooved-tip designs effectively increase the pressure within the \ac{ptv} core, thereby reducing the pressure drop and lowering the risk of tip cavitation. 

Among the four grooved-tip designs, all demonstrate comparable velocity deficits in the \ac{ptv} region, as evidenced in Figure~\ref{fig:streamwisepiv_Vmag}.
This consistency is attributed to their similar effective \ac{2d} permeability, which governs the interaction between the grooves and the flow.
Overall, considering both the amplitude of velocity deficits (Figure~\ref{fig:line_vel_mag}) and the lift-drag performance (Figure~\ref{fig:line_CL_CD}), the 25G-C-CN design appears to be the most promising configuration among those tested.
Consequently, this design has been selected for a more detailed comparative analysis against the baseline case in subsequent sections.

\subsection{Vortex pattern}\label{sec:3.3}
\begin{figure*}[tbp!]
    \centering
    \includegraphics[width=8.358cm]{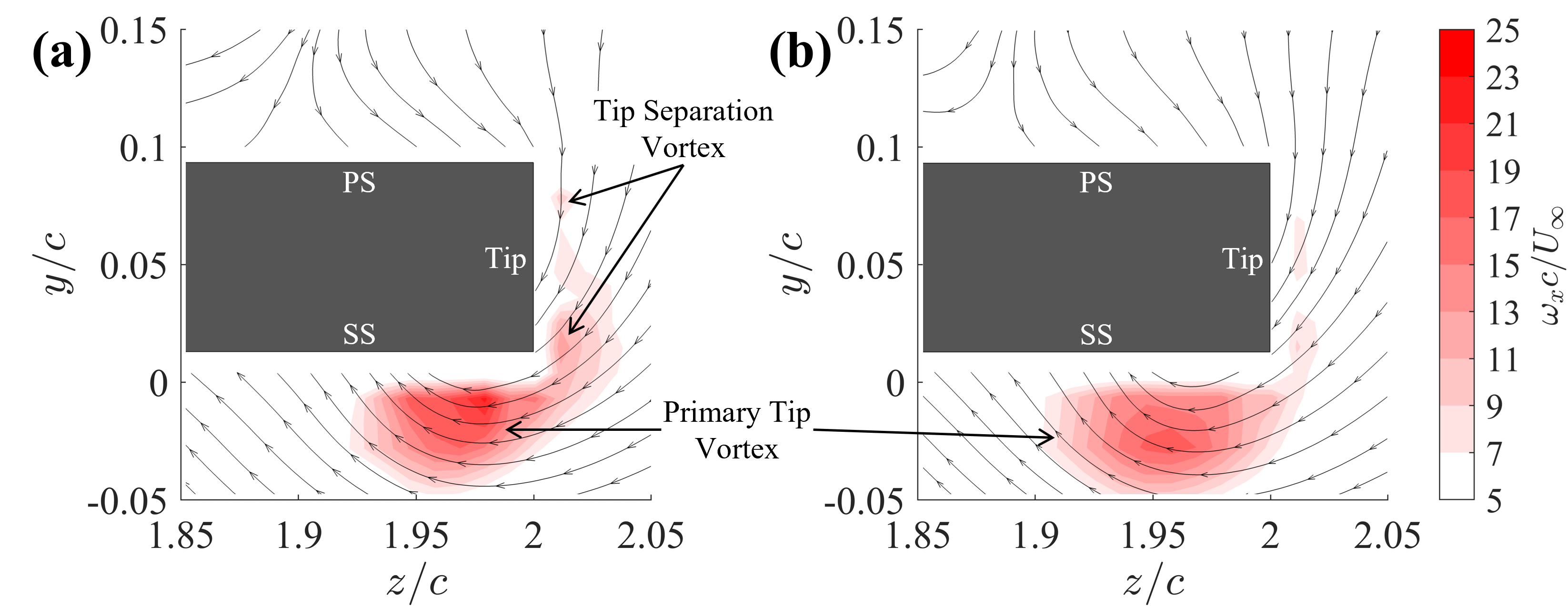}
    \caption{Time-averaged vorticity contours superimposed with time-averaged streamlines for cross-flow plane at $x/c = 0.7$ for (a) baseline, (b) 25G-C-CN.}
    \label{fig:crossflowpiv_vorticity}
\end{figure*}

This section aims to investigate how the grooved-tip designs fundamentally alter the vortex structures, consequently affect the intensity of the tip vortices.
The key finding is that the grooved-tip design significantly suppresses the \ac{tsv} and leads to a weaker, more diffused \ac{ptv}. Figure~\ref{fig:crossflowpiv_vorticity} presents a comparison of the time-averaged streamwise vorticity, $\omega_x = \partial w / \partial y - \partial v / \partial z$, and streamlines between the baseline case and the 25G-C-CN design at the cross-flow plane of $x/c=0.7$.
In the baseline case (Figure~\ref{fig:crossflowpiv_vorticity}(a)), two distinct components of tip vortices are identified: the \ac{ptv}, located just below the suction side, and the \ac{tsv} which is observed to grow in size from the pressure side to the suction side.
There is also evident entangling and mixing of the \ac{ptv} and \ac{tsv} in the baseline configuration.
In contrast, Figure~\ref{fig:crossflowpiv_vorticity}(b) demonstrate a significant reduction in the magnitude of vorticity for the \ac{tsv} due to the groove-tip design, indicating that the grooves are highly effective in weakening this vortex.
Furthermore, the grooved-tip design results in a noticeable reduction in vorticity magnitude of the \ac{ptv}, accompanied by an enlarged vortex size, which is a clear sign of enhanced vortex diffusion.

\begin{figure*}[tbp!]
    \centering
    \includegraphics[width=4.73cm]{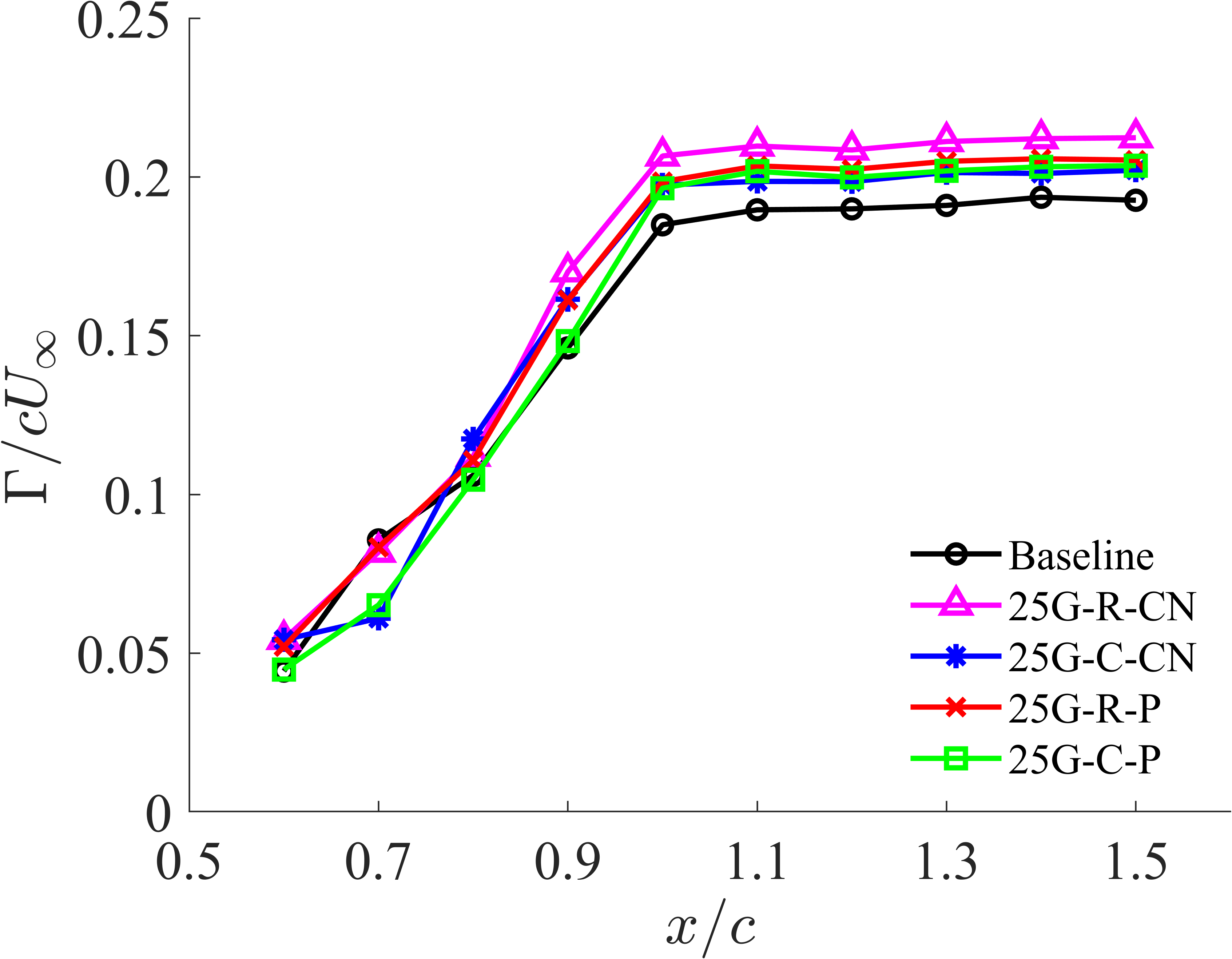}
    \caption{Comparison of the non-dimensional vortex circulation $\Gamma/cU_\infty$ on each cross-section along the vortex core trajectory between baseline and grooved tips.}
    \label{fig:line_circulation}
\end{figure*}

To further understand the underlying physics of how grooved-tip treatments weaken tip vortices and mitigate the associated pressure drop, we examined key vortex parameter such as vortex circulation.
Figure~\ref{fig:line_circulation} presents the non-dimensional vortex circulation, $\Gamma/cU_\infty$, obtained by integrating the streamwise vorticity $\omega_x$ at each streamwise station from cross-flow \ac{piv} measurements.
A threshold was applied to include only the positive vorticity, accounting for the \ac{ptv}'s contribution along its trajectory while excluding negative vorticity from the \ac{tev}.
Upstream of the trailing edge, the circulation grows along the streamwise direction with little variation in its rate, and it tends to plateau beyond the trailing edge.
Notably, the circulation values for each groove-tip design and the baseline are close at each cross-section, especially upstream of the trailing edge.
This suggests that the total circulation generated by the tip vortices remains largely unaffected by the grooves, which aligns with the slight impact of the grooved-tip designs on the lift and drag performance, as shown in Figure~\ref{fig:line_CL_CD}.

\subsection{Vortex strength mitigation}\label{sec:3.4}
This section quantifies the attenuation of the \ac{ptv} strength due to the grooved-tip design.
In the preceding section, Figure~\ref{fig:crossflowpiv_vorticity} suggested that the grooved-tip design leads to both an attenuation in the vorticity and an enlargement of vortex size.
To qualitatively assess the vortex strength, the non-dimensional vortex swirling strength ($\lambda_{c} c / U_\infty$), is employed.
This parameter, defined as the imaginary part of the complex eigenvalues of the velocity gradient tensor \citep{chong1990,zhou1999}, is effective in identifying the vortex core regions and representing vortex intensity \citep{Jeong1995,chakraborty2005}.

\begin{figure*}[tbp!]
    \centering
    \includegraphics[width=8.4cm]{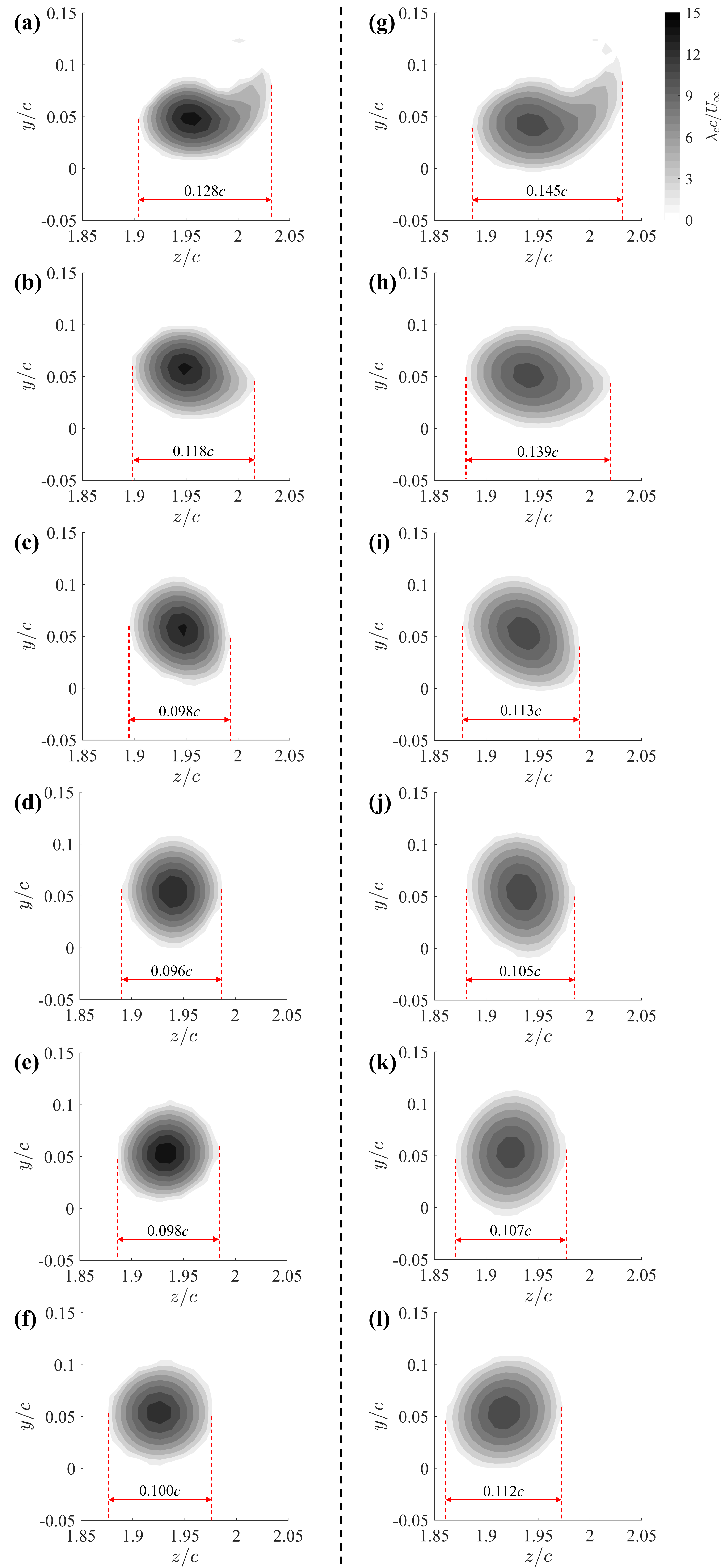}
    \caption{Contours of non-dimensional swirling strength $\lambda_{c} c / U_\infty$ on each cross-section along the vortex core trajectory downstream of the trailing edge for baseline tip (a--f) and 25G-C-CN grooved tip (g--l). (a, g) $x/c = 1$; (b, h) $x/c = 1.1$; (c, i) $x/c = 1.2$; (d, j) $x/c = 1.3$; (e, k) $x/c = 1.4$; (f, l) $x/c = 1.5$.}
    \label{fig:crossflowpiv}
\end{figure*}

Figure~\ref{fig:crossflowpiv} displays contours of the non-dimensional vortex swirling strength at successive cross-sections downstream of the trailing edge for the baseline case (Figures~\ref{fig:crossflowpiv}(a)--(f)) and the 25G-C-CN design (Figures~\ref{fig:crossflowpiv}(g)--(l)).
At the $x/c=1$ plane, both tip designs show the \ac{ptv} with a small ``tail,'' indicative of the tip vortex's early-stage rolling-up process.
The size of this ``tail'' is more prominent in the 25G-C-CN design, potentially due to the shearing effect from the groove channel flows.
As the vortex convects downstream, it becomes more circular and axisymmetric.
A well-defined vortex core is visible in each cross-section for both cases, confirming that the tip vortex remains coherent in the near-wake region from $x/c=1$ to at least $x/c=1.5$ (which is the most downstream plane measured).
As anticipated, the strongest swirling appears at the vortex core centre, with weaker rotation observed at increasing radial distance from the centre.
The relatively high peak magnitudes in each slice suggests the absence of a drastic vortex breakdown as the vortex convects downstream.
A gradual shift of the vortex centre towards the wing root direction is observed.

It is also noticed that Figures~\ref{fig:crossflowpiv}(g)--(l) clearly demonstrate the \ac{ptv} core of 25G-C-CN grooved-tip design exhibiting larger dimensions and lower maximum swirling strength at the vortex centre compared to the baseline case (Figures~\ref{fig:crossflowpiv}(a)--(f)).
This visual evidence strongly suggests that the grooved-tip treatment leads to a weaker \ac{ptv}, which has significantly enlarged dimensions.
The increased vortex dimensions suggest that the energy contained in the vortex is dissipated over a larger area, given Figure~\ref{fig:line_circulation} has shown that the circulation for different tip designs remain close, which will contribute to the reduced pressure drops.

The inception of cavitation is directly linked to local pressure drops, for example, within the core of a swirling vortices.
Therefore, understanding the pressure change at the vortex core along the \ac{ptv} trajectory is critical for assessing the mitigation effect of the grooved-tip design on cavitation risks.
Due to the limitations of the stereoscopic \ac{piv} setup, which lacks the third dimension of velocity vectors and their derivatives, direct pressure measurement for highly three-dimensional flow features like tip vortices in not feasible.
Consequently, a reduced-order model was adopted to estimate the pressure change within the \ac{ptv} core relative to the free stream, as detailed below.

\subsubsection{Reduced-order model for pressure estimation}
\begin{figure*}[tbp!]
    \centering
    \includegraphics[width=8cm]{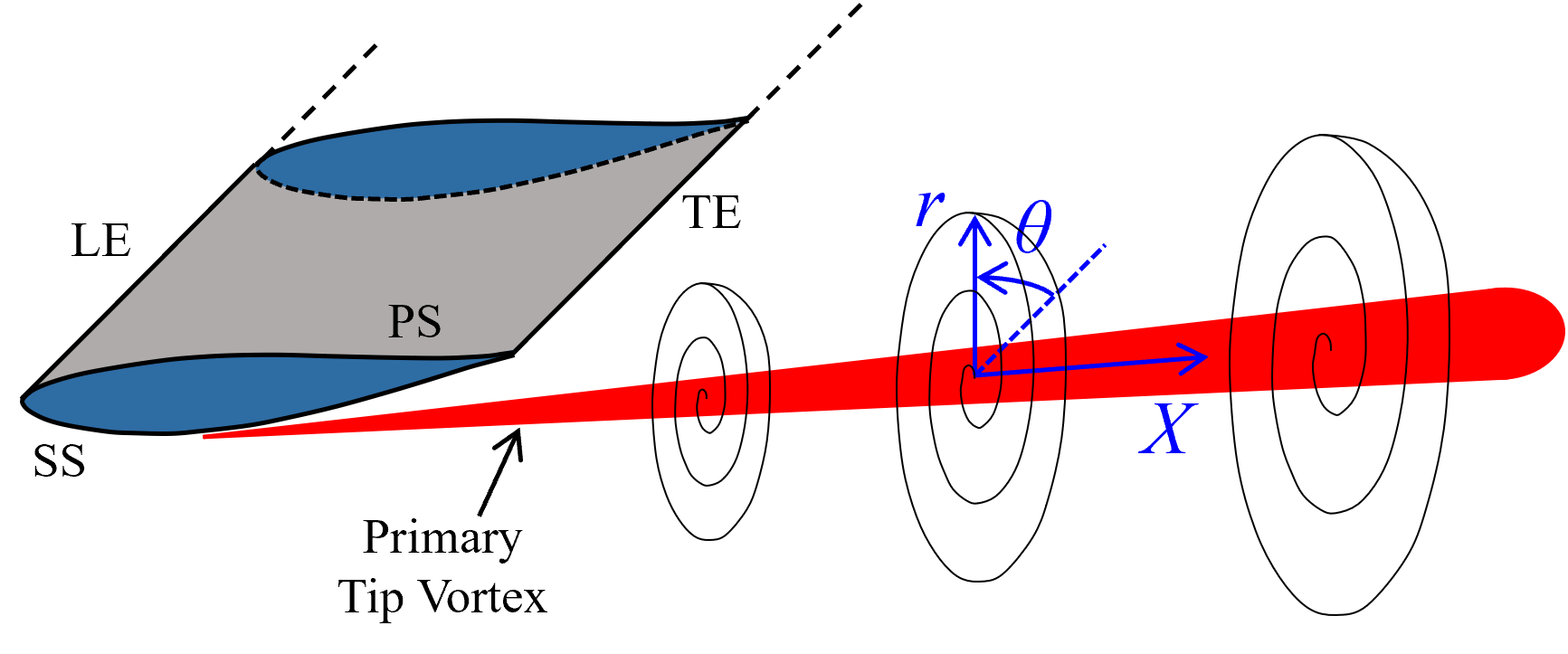}
    \caption{Cylindrical polar coordinate system used for simplified pressure analysis model; not to scale. LE: leading edge; TE: trailing edge.}
    \label{fig:polar}
\end{figure*}

To assess the local pressure change within the tip vortex core, an analytical model based on a cylindrical polar coordinate system O$(r, \,\theta, \,X)$ with its origin centred at the vortex centre and velocity components $(v_r, \,v_\theta, \,v_X)$ was adopted.
Here, $X$ represents the tip vortex axial direction, which is almost parallel with the streamwise direction.
Radial velocity $v_r$ and swirl velocity $v_\theta$ approach zero in the free stream far from the tip vortex, while axial velocity $v_X$ approaches $U_\infty$ at infinity.
The cylindrical polar coordinate system is illustrated in Figure~\ref{fig:polar}.
For a steady, incompressible, axisymmetric vortex, the equation of motion over a cross-section of the \ac{ptv} at $X=\mathrm{constant}$ yields the pressure defect on the axis as \citep{morton1969}
\begin{equation}
    p_\infty - p_0(X) = \rho \int_{0}^{\infty} \frac{v_\theta^2}{r}\,\mathrm{d}r,
    \label{eq:2}
\end{equation}
where $p$ is pressure, $r$ is radial distance from the axis, and $\rho$ is the fluid density.
Assuming the radial velocity of the vortex core is zero, the main driver of the radial pressure gradient simplifies to $v_\theta^2/r$, establishing a direct link between the swirl intensity and radial pressure gradient.

For simplicity, the vortex is considered two-dimensional on a \ac{2d} measurement plane, i.e., neglecting the axial velocity, and further making the assumption that the vortex behaves like a Rankine vortex.
Similar approaches have been adopted by \citep{dreyer2015} to represent tip vortices using simplified vortex models for \ac{2d} data, and inferring the minimum pressure coefficient within the tip vortex core \citep{billet1981,boulon1999,franc2005}.
In a \ac{2d} Rankine vortex model, the velocity profile in cylindrical polar coordinates ($r, \theta, X$) is given by
\begin{equation}
    v_\theta (r) =
    \begin{cases}
        \Omega r, & \text{for } r \leq r_c \quad \text{(solid-body rotation)} \\
        \frac{\Gamma}{2\pi r}, & \text{for } r > r_c \quad \text{(potential vortex)}
    \end{cases}
    \label{eq:3}
\end{equation}
where $\Gamma=$ is the circulation of the vortex, $r_c$ is the vortex viscous core radius, and $\Omega=\Gamma/2\pi r_c^2$ is the constant angular velocity of the rigid core.
Substituting Equation~(\ref{eq:3}) into Equation~(\ref{eq:2}) and integrating from the vortex centre ($r=0$, where the pressure drop is maximum), to infinity, yields
\begin{align}
    \begin{split}
        p_\infty    &   = p_{r_c} + \rho \frac{\Gamma^2}{8 \pi r_c^2}, \quad \text{and} \\
        p_0         &   = p_{r_c} - \frac{1}{2} \rho \Omega^2 r_c^2.
    \end{split}
    \label{eq:4}
\end{align}
Hence, the pressure drop at the vortex core centre is
\begin{align}
    \begin{split}
        \Delta p_\mathrm{max}   &   = p_\infty - p_0 \\
                                & = \rho \frac{\Gamma^2}{8 \pi^2 r_c^2} + \frac{1}{2} \rho \Omega^2 r_c^2.
    \end{split}
    \label{eq:5}
\end{align}
Given the circulation of a Rankine vortex is $\Gamma=2 \pi \Omega r_c^2$, Equation~\ref{eq:5} simplifies to
\begin{equation}
    \Delta p_\mathrm{max} = \rho \Omega^2 r_c^2.
\end{equation}

For a purely rotating flow, the vortex swirling strength $\lambda_{c}$ can be shown equal to the angular velocity $\Omega$ of the rotating core.
Therefore, based on this idealised model, the pressure drop is directly proportional to the square of the swirling strength
\begin{equation}
    \Delta p \propto \lambda_{c}^2.
    \label{eq:6}
\end{equation}

\subsubsection{Pressure mitigation by the grooved-tip}
The reduced-order model demonstrates that the pressure drop within the vortex core is directly proportional to the square of the vortex swirling strength.
Consequently, the vortex swirling strength contours presented in Figure~\ref{fig:crossflowpiv} suggest a significant mitigation in the pressure drop inside the tip vortex core, as the maximum vortex swirling strength is notably reduced by the grooved-tip design.
While this simplified vortex model neglects the axial velocity, Section~\ref{sec:3.2} also showed a significant deficit in the axial velocity caused by the grooved tip, suggesting an even greater mitigation in the pressure drop within the \ac{ptv} that indicated by the swirling strength alone.

\begin{figure*}[tbp!]
    \centering
    \includegraphics[width=5.06cm]{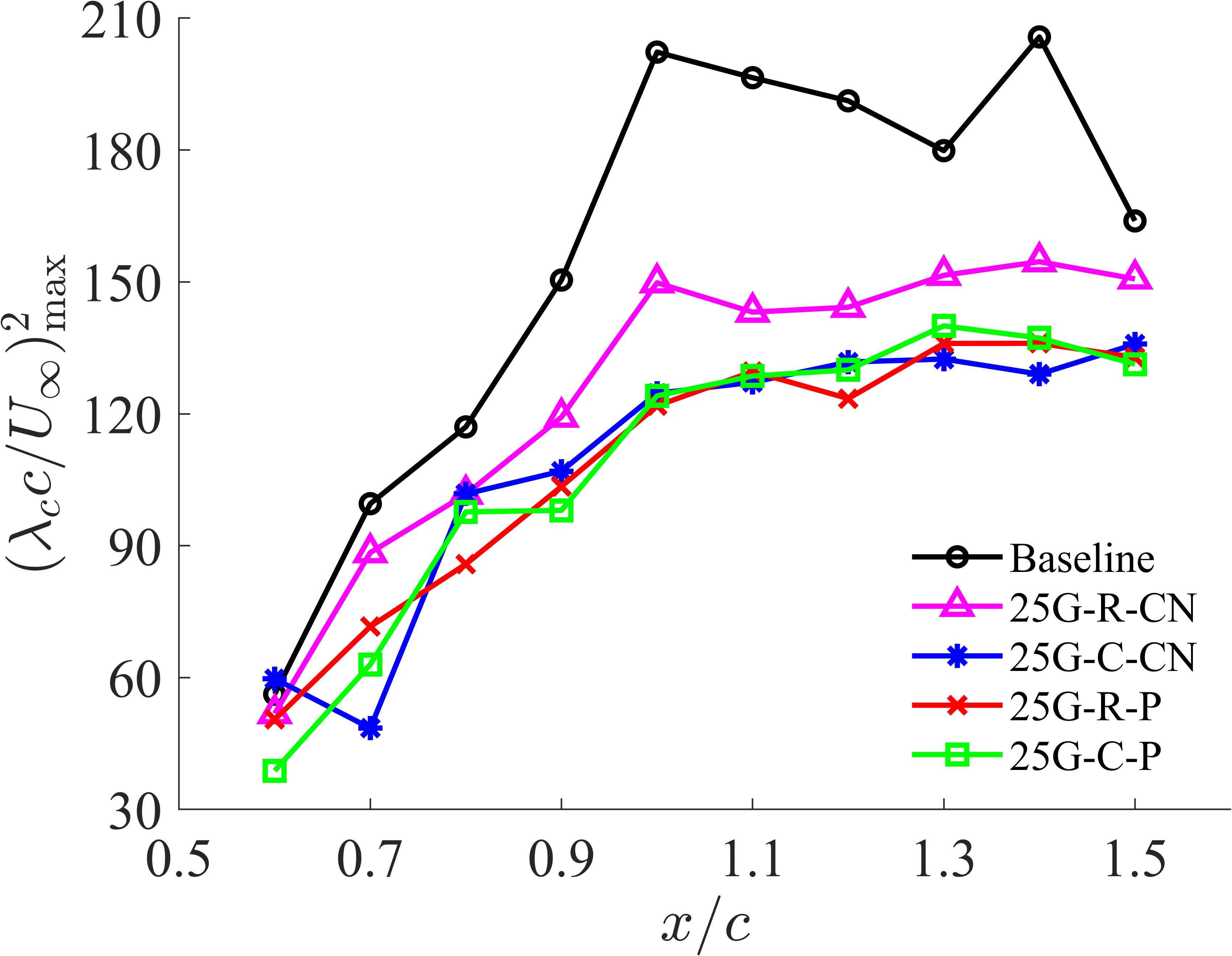}
    \caption{Comparison of maximum vortex swirling strength $\lambda_{c} c / U_\infty$ on cross-sections along the vortex core trajectory before the trailing edge between baseline and grooved tips.}
    \label{fig:line_swirling_strength}
\end{figure*}

Figure~\ref{fig:line_swirling_strength} presents the variation of the maximum swirling strength $\lambda_{c}$ (normalised by the chord length $c$ and free stream velocity $U_\infty$) as a function of streamwise distance $x/c$ for the baseline and the four grooved-tip designs.
Near the mid-chord, maximum $\lambda_{c}$ values are relatively low and comparable for the baseline and grooved tips.
As the \ac{ptv} develops towards the trailing edge, the maximum swirling strength increases, indicating that the vortical structures intensify along the tip chord.
This growth is driven by the continuous entrainment of tip flows due to the pressure difference between the pressure and suction sides of the wing.
Beyond the trailing edge, the baseline case shows to decay gradually, signifying vortex diffusion and dissipation as the vortex convects downstream.
In contrast, the grooved-tip cases level off, suggesting that the diffusion and dissipation are expedited by the groove structures even before the trailing edge.

Consistently, the baseline case exhibits the highest swirling strength at each cross-section, confirming that the unmodified tip generates the strongest vortex.
The 25G-R-CN case achieved moderate mitigation of vortex strength, remaining below the baseline curve but relatively high compared to other grooved-tip designs; a reduction of 27\% was observed at $x/c=1.1$.
Importantly, the remaining three designs, including 25G-C-CN, 25G-R-P and 25G-C-P, demonstrate similar and more substantial reductions in vortex strength.
The most significant pressure-drop mitigation, nearly 40\%, is always achieved at $x/c=1.0$ by these three grooved-tip designs.
This data strongly supports the effectiveness of the proposed grooved-tip designs in mitigating tip vortex-induced pressure drops and, consequently, cavitation risk.

\section{Conclusions}\label{conclusions}
The present work investigates the effects and underlying physics of introducing tip permeability, realised by a novel multiple-groove design at the tip, on the characteristics of tip vortices.
The experiments were performed on a NACA~63-415 wing in a water tunnel at a chord-based Reynolds number of $3.22\times 10^4$, with a semi-aspect ratio of 2.
Lift and drag were measured across a wide range of angles of attack, and flow fields near the wing tip and along the \ac{ptv} trajectory were measured using \ac{piv} at $\alpha=6^\circ$.

The proposed groove-tip designs demonstrate negligible differences in lift and drag coefficients within the typical range of angles of attack, compared to the baseline wing with an impermeable tip.
Notably, these designs also exhibit a moderate delay in stall for the aspect ratio tested ($s\AR=2$), which will be further investigated in future work.

Streamwise velocity measurements near the wing tip ($z/c=1.95$) consistently showed a significant velocity deficit within the \ac{ptv} region for all grooved-tip designs tested, compared to the baseline case.
This velocity reduction, predominantly attributed to the changes in the streamwise component, suggests a modification of the tip vortex structure, which was caused by the impingement and mixing of the groove jets with the \ac{ptv}.

Analysis of the cross-flow plane at $x/c=0.7$ revealed that the baseline case exhibited entanglement between the \ac{tsv} and the concentrated \ac{ptv}, which contributed to intensified tip vortex structures.
In contrast, the grooved-tip design almost eliminated the \ac{tsv} and resulted in a more diffused \ac{ptv}, evidenced by the weakened vorticity and enlarged vortex dimensions.
Vortex parameter analysis indicated only a minimal change in the vortex circulation for cross-sections of the \ac{ptv} before the trailing edge, irrespective of the tip design.
However, the grooved-tip structures notably increased the size of the \ac{ptv}.

A reduced-order model was employed to qualitatively compare vortex strengths, addressing the limitations of experimental measurements.
This model established that for a \ac{2d} Rankine vortex, the pressure drop at the vortex core is proportional to the square of the vortex swirling strength.
Consequently, all four grooved-tip designs showed reduced maximum swirling strength along the \ac{ptv} trajectory compared to the baseline case.
This indicates that the proposed designs are effective in modifying the flow field within the tip vortex, resulting in a pressure drop reduction of up to 40\% compared to the baseline case at the tested Reynolds number.
Such a reduction can significantly mitigate cavitation risks induced by tip vortices in wing- and blade-based systems.

The demonstrated passive flow control technology, achieved through grooved-tip treatment, offers a promising approach for modifying tip vortex characteristics in various applications, including those of hydrofoils on marine vessels and underwater vehicles, as well as turbines, propellers and pumps.
Furthermore, the development of advanced permeable structures holds significant potential for noise suppression in wind turbines and aerial vehicles.

Constrained by the current experimental setup and capabilities, the two-dimensional data are insufficient to fully understand these highly three-dimensional flow features as well as internal flow fields inside the grooves.
Further investigation could benefit from more advanced volumetric measurement techniques, such as the time-resolved three-dimensional Lagrangian Particle Tracking Velocimetry or tomographic \ac{piv}, coupled with pressure reconstruction algorithms.

\backmatter
\bmhead{Author contributions}
All authors collaboratively designed and conceived the study.
JT -- designed, prepared and performed the experiments, processed and analysed the experimental data, and wrote the first draft of this manuscript.
SO -- advised on the experiments and data analysis, and edited the manuscript.
IMV -- co-supervised the project, advised on the experiments and simulations, and edited the manuscript.
YL -- supervised the project, performed the 2D simulations, advised on the experiments and data analysis, secured funding, and edited the manuscript.

\bmhead{Acknowledgements}
This work has been supported by the Royal Commission for the Exhibition of 1851 Brunel Fellowship (YL), the UK Engineering and Physical Science Research Council (EPSRC) funded Supergen ORE Hub [EP/Y016297/1] Flexible Fund, Standard Proposal [EP/V009443/1] and IAA Innovation Competition Award [EPSRC IAA PV111], the Royal Society ISPF-International Collaboration Award [ICA/R1/231053] and International Exchanges [WT13324749], and Japan Society for the Promotion of Science KAKENHI [JP24K17204]. The experiments were performed in the School of Engineering, University of Edinburgh.

\bmhead{Data Availability} The data that support the findings of the study are available from the corresponding author upon reasonable request.

\section*{Declarations}
\bmhead{Completing Interests}
The authors have no conflict of interest to report.

\bmhead{Ethical Approval}
Not applicable.

\bibliography{Manuscript}

\begin{appendices}
\section{Force measurement uncertainties} \label{app:uncertainty}
The force coefficient $C_F$, which can either be the lift or drag coefficient, is written as
\begin{equation}
    C_F = \frac{F}{0.5\rho U_\infty^2S}.
\end{equation}
Therefore, the total bias of $C_F$ is written as
\begin{align}
    B_T(C_F) = \Bigg[ 
        \left( \frac{\partial C_F}{\partial F} \, B(F) \right)^2 +
        \left( \frac{\partial C_F}{\partial \rho} \, B(\rho) \right)^2 \notag \\
        + \left( \frac{\partial C_F}{\partial U_\infty} \, B(U_\infty) \right)^2 +
        \left( \frac{\partial C_F}{\partial S} \, B(S) \right)^2
    \Bigg]^{1/2}
    \label{eq:a2}
\end{align}
where $B(F)$, $B(\rho)$, $B(U_\infty)$ and $B(S)$ are the bias limits of the force, fluid density, free stream velocity and area, respectively. Each term is estimated as below.

$B(F)$: The half-wing model was rigidly attached to the load cell in the experiments, so that the inclination of the load cell is dependent on the wing's angle of attack $\alpha$. Axis rotation of the measured force ($F_x$, $F_y$) in the load cell's reference frame to the global reference frame yields
\begin{align}
    \begin{split}
        L   & = F_x\sin(\alpha)+F_y\cos(\alpha), \quad \text{and} \\
        D   & = F_x\cos(\alpha)-F_y\sin(\alpha).
    \end{split}
\end{align}
As per calibration report of the ATI Industrial Automation Nano17 IP68 load cell, the bias limits in $F_x$ and $F_y$ force components are 0.07\% and 0.06\% of the sensing range (25~N). Therefore, the bias limits of the measured forces are $B(F_x)=0.0183$~N and $B(F_y) = 0.0158$~N, respectively. The load cell and the half-wing model were rigidly connected to a beam above the free water surface. The beam was parallel with the load cell's horizontal axis as well as the wing's chord. Therefore, the angle of attack of the wing $\alpha$ is equivalent to the rotation angle $\eta$ of the beam. The rotation angle $\eta$ was measured by a digital inclinometer with an accuracy of $\pm0.025^\circ$, so that $B(\alpha) = B(\eta) = 0.025^\circ$. The lift and drag forces each has three uncorrelated uncertainties, namely $B(F_x)$, $B(F_y)$ and $B(\alpha)$. The bias limits $B(L)$ and $B(D)$ for the lift and drag can be written as
\begin{align}
\begin{split}
    B(L) &= \Big[ 
        \left( \sin(\alpha) \, B(F_x) \right)^2 +
        \left( \cos(\alpha) \, B(F_y) \right)^2 \\
        &\quad + \left( \left( F_x \cos(\alpha) - F_y \sin(\alpha) \right) B(\alpha) \right)^2 
    \Big]^{1/2}, \quad \text{and} \\
    B(D) &= \Big[ 
        \left( \cos(\alpha) \, B(F_x) \right)^2 +
        \left( \sin(\alpha) \, B(F_y) \right)^2 \\
        &\quad + \left( \left( F_x \sin(\alpha) + F_y \cos(\alpha) \right) B(\alpha) \right)^2 
    \Big]^{1/2},
\end{split}
\label{eq:lift_drag_uncertainty}
\end{align}
respectively.

$B(\rho)$: Before and after each set of force measurement, the variation of the temperature $T$ measured by a handheld digital thermometer (RS PRO 206-3738) with an accuracy of $1^\circ\mathrm{C}$ was between $0.1^\circ\mathrm{C}$ and $0.5^\circ\mathrm{C}$. Consequently, the minimum and maximum bias limits of the temperature measurement are $B_\mathrm{min}(T)=1.1^\circ\mathrm{C}$ and $B_\mathrm{max}(T)=1.5^\circ\mathrm{C}$. The bias limit of water density as a function of temperature is \citep{ITTC2011}
\begin{equation}
    B(\rho) = \left|\frac{\partial \rho}{\partial T}\right|B(T),
    \label{eq:a4}
\end{equation}
where the sensitivity coefficient is
\begin{equation}
     \left|\frac{\partial \rho}{\partial T}\right| = \left|0.0552-0.0154T+0.000120T^2\right|.
     \label{eq:a5}
\end{equation}
Equations~\ref{eq:a4} and \ref{eq:a5} yields the following bias limits: $B_\mathrm{min}(\rho) = 0.2171$~kg~m$^{-3}$ and $B_\mathrm{max}(\rho) = 0.2960$~kg~m$^{-3}$, respectively. The density change due to the seeding particles for flow visualisation is neglected.

$B(U_\infty)$: The free stream velocity of the flow was not measured during the measurements to avoid flow disruptions from the measurement device. However, the water tunnel was calibrated before the experiments by a Vectrino \ac{adv}. The turbulence intensity in the $x$, $y$ and $z$ directions are $0.0480U_\infty$, $0.0426U_\infty$ and $0.0267U_\infty$, respectively. Therefore, the overall bias limit of the free stream velocity is estimated as $0.0695U_\infty$, giving $B(U_\infty)=0.021$\mps. This is consistent with previously reported experiments conducted in the same water tunnel \citep{arredondo2023}.

$B(S)$: The dimensions of the wing was measured after sanding, so that the accuracy of the 3D printer is not used here. The ruler has resolution of 1~mm and an accuracy of $\pm0.5$~mm. Therefore, the bias limit for the area, $S=bc$, considering the bidimensionality of the area, is $B(S)=2.5\times10^{-7}$~m for all wings. 

\begin{figure*}[tbp!]
    \centering
    \includegraphics[width=8cm]{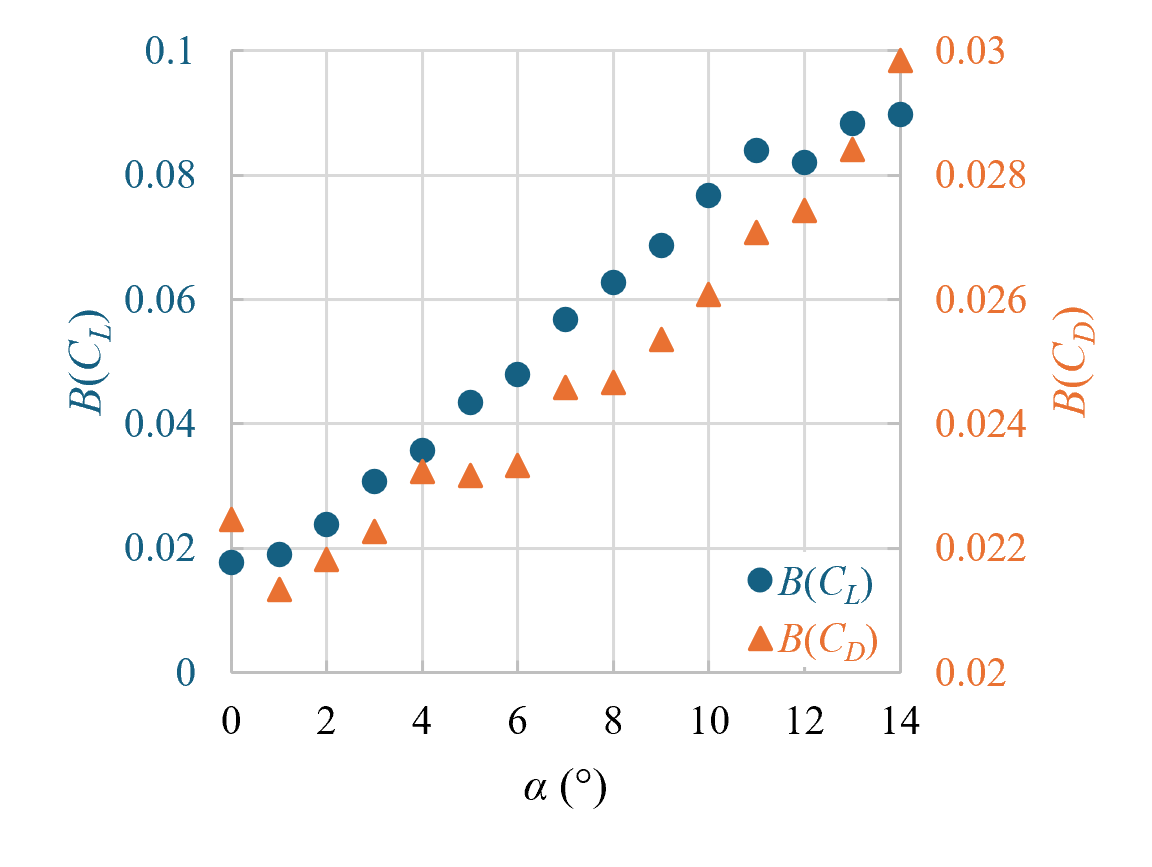}
    \caption{Bias limits of force coefficients as a function of angle of attack $\alpha$.}
    \label{fig:uncertainty}
\end{figure*}

The bias limits for all inputs are summarised in Table~\ref{tab:uncertainty}, together with the maximum total bias limit for the lift $B_{T_\mathrm{max}}(C_L)$ and drag $B_{T_\mathrm{max}}(C_D)$ coefficients at $Re_c=3.22\times10^{-4}$ computed using Equation~\ref{eq:a2}. The bias limits $B(C_L)$ and $B(C_D)$, as functions of $\alpha$ are shown in Figure~\ref{fig:uncertainty}.

\begin{table}[tbp!]
    \centering
    \caption{Summary of bias limits for force measurement.}
    \label{tab:uncertainty}
    \begin{tabular}{m{7.5em} m{7.5em} m{7.5em}}
        \hline
        \textbf{Variable}   & \textbf{Magnitude} & \textbf{Unit} \\
        \hline
        $B(L)$  & 0.0159 & [N] \\
        $B(D)$  & 0.0204 & [N] \\
        $B(\rho)$  & 0.2960 & [kg~m$^{-3}$] \\
        $B(U_\infty)$ & 0.021 & [m~s$^{-1}$]  \\
        $B(S)$  & $2.5\times10^{-7}$ & [m$^2$] \\
        $B_{T_\mathrm{max}}(C_L)$  & 0.090 & $-$ \\
        $B_{T_\mathrm{max}}(C_D)$ & 0.030 & $-$  \\
        \hline
    \end{tabular}
\end{table}

\section{Configuration of grooves through 2D permeability estimation} \label{app:2dpermeability}
\begin{figure*}[tbp!]
    \centering
    \includegraphics[width=86mm]{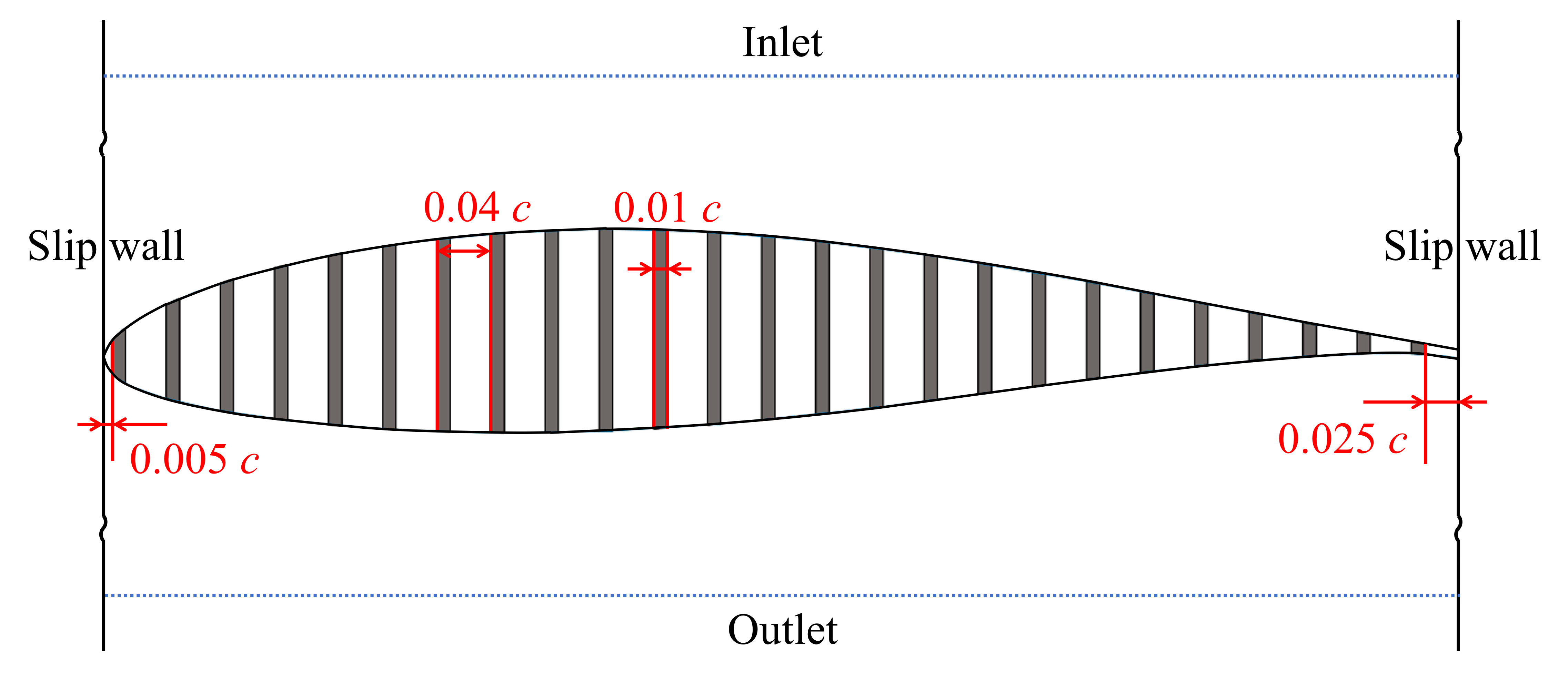}
    \caption{Schematic of 2D channel simulation with a multi-groove configuration.}
    \label{fig_2DChannel}
\end{figure*}

To inform the practical design of multi-groove configurations, a two-dimensional (2D) channel simulation is carried out.
As shown in Figure~\ref{fig_2DChannel}, the setup features a foil embedded within the channel that replicates the tip profile of a turbine blade.
Two simulation scenarios are considered: (i) the foil is modelled as a porous medium with uniform permeability, characterised by the non-dimensional Darcy number ($Da$); and (ii) the foil incorporates multiple grooves that permit fluid passage, while the remaining sections remain impermeable.
The groove configuration is considered to exhibit equivalent 2D (chord-normal) permeability to the porous tip when both cases yield the same pressure drop between the inlet and outlet of the channel.
Based on prior work identifying the optimal permeability range as $Da = 10^{-6}$ to $10^{-5}$ (see \cite{liu2024arxiv}), the corresponding groove geometries are selected to fall within this effective permeability range.

\end{appendices}

\end{document}